\documentclass[aps,prb,twocolumn,superscriptaddress]{revtex4-2}
\usepackage{graphicx}
\usepackage{epsfig}
\usepackage{color}
\usepackage{stackrel}
\usepackage{amsfonts}
\usepackage{wrapfig}
\usepackage{pstricks}
\usepackage{pst-node}
\usepackage{bm}
\usepackage{dcolumn}
\usepackage{multirow}
\usepackage{epstopdf}
\usepackage{subfigure}
\usepackage{amssymb}
\usepackage{amsmath}
\usepackage{animate}
\usepackage{commath}
\usepackage{verbatim}
\usepackage{booktabs}
\usepackage[para]{threeparttable}
\usepackage{etoolbox}
\usepackage{lipsum}
\usepackage{marvosym}
\usepackage{wasysym}
\usepackage{kotex}
\usepackage{hyperref}

\begin{document}
\title{Excitons in moir\'{e} superlattices with disordered electrons}

\author{Junghwan Kim}
\altaffiliation{jkim392@ur.rochester.edu}
\affiliation{Department of Electrical and Computer Engineering, University of Rochester, Rochester, New York 14627, USA}

\author{Dinh~Van~Tuan}
\altaffiliation{vdinh@ur.rochester.edu}
\affiliation{Department of Electrical and Computer Engineering, University of Rochester, Rochester, New York 14627, USA}

\author{Hanan~Dery}
\altaffiliation{hanan.dery@rochester.edu}
\affiliation{Department of Electrical and Computer Engineering, University of Rochester, Rochester, New York 14627, USA}
\affiliation{Department of Physics and Astronomy, University of Rochester, Rochester, New York 14627, USA}

\begin{abstract}
Moir\'{e} superlattices in transition metal dichalcogenides (TMDs) heterobilayers exhibit various correlated insulating states driven by long-range Coulomb interactions, and these states crucially alter exciton resonances, particularly at fractional fillings. We revisit a theoretical framework to investigate the doping dependence of exciton spectra by extending hydrogenic exciton wavefunctions, systematically analyzing how the 1$s$, 2$s$, and 3$s$ Rydberg states respond to moiré-induced mixing of $s$- and $p$-type orbitals. Notably, while the 1$s$ state remains relatively robust against doping, higher Rydberg excitons show strong redshifts and oscillator-strength quenching near specific fractional fillings. We incorporate both defect-induced quasi-ordering and thermal fluctuations to capture realistic device conditions, employing a large supercell approach. By selectively randomizing a subset of electrons or utilizing classical Monte Carlo simulations, we present direct calculations of exciton spectra under varying defect densities and temperatures. Our results emphasize how even moderate disorder or finite temperature can partially or completely suppress characteristic moir\'{e} exciton physics. Especially, we show how the 2$s$ exciton states respond to the phase transition in correlated electron states. This comprehensive picture not only clarifies recent experimental observations but also provides a framework to guide the design of moir\'{e}-based optoelectronic devices.
\end{abstract}

\maketitle

\section{Introduction}
Two-dimensional (2D) TMDs such as MoS$_2$, WS$_2$, MoSe$_2$, and WSe$_2$ host tightly bound excitons with binding energies on the order of hundreds of meV \cite{MuellerMalic,Raja2017,ZhangReview,Comp_Wang, Mak2010, He2014, Chernikov2014, Ugeda2014,our_PRB_2018}. In the monolayer limit, the ultra-strong Coulomb interactions give rise not only to bright intralayer excitons but also to a rich set of many-body quasiparticles~\cite{basic_Selig,RefS1,RefS2,RefS3,RefS4,RefS5,RefS6,RefS7,RefS8,RefS9,RefS10,RefS11,our_PRB_2020,our_PRL_2022, our_PRB_2022, upconv_Jones2016, valleycoherence_Jones2013, valleyphonon_He2020}. When two different TMDs monolayers are stacked with a small twist angle or slight lattice mismatch, they form a moir\'{e} superlattice with a long-range periodic potential~\cite{Moire1_Tang,Moire2_Seyler,WuHubbard2018,ZhangFlat2020,MakReview2022,TranMoireExcitons2019}. At fractional fillings of the moir\'{e} minibands, strong correlation effects can emerge, including Mott-insulating and Wigner-crystal states~\cite{Fractional1_Regan,Fractional2_Shimazaki,Fractional3_Wang,Li2021, Lian2023, Xu2020,moireexciton_Wang2023}. These intriguing strong correlation phenomena have prompted extensive theoretical research, and a variety of fascinating physical effects have also been observed in experiments~\cite{MoireExcitons_Jin,Zhang2021, Xiong2023, Campbell2022, moire_Zhang2022, moire_Gu2022, moire_Zhang2022b, moire_Zeng2023, moire_Wu2018, moire_Tang2021, moire_Zhang2021_U4, moire_Li2021, moire_Anderson2024, moire_Wang2025, BenMhenni2024}.

Recent studies experimentally identify that the 2$s$ exciton is particularly sensitive to fractional fillings and electron correlations in moir\'{e} heterobilayers~\cite{Xu2020,Gu2024,Hu2023}. 
{ In the experimental architecture that our theory models, this sensitivity is probed optically in a separate `sensor' monolayer, whose excitons are coupled to the moiré layer via the long-range Coulomb interaction.
}
Its resonance may vanish, shift, or be strongly quenched depending on the doping level. In contrast, the 1$s$ exciton often remains relatively robust, showing minimal shifts even under variable doping. Our previous study demonstrated that the pronounced sensitivity of the 2$s$ exciton to fractional fillings arises from strong 2$s$--2$p$ hybridization mediated by the moir\'{e}-extended wavefunction~\cite{our_previous_work}. 
{
The moir\'{e}-extended wavefunction refers specifically to the exciton's center-of-mass wavefunction forming a Bloch state that propagates through the periodic potential induced by the charge-ordered Wigner crystal. This extended state, responsible for the sharp absorption resonances, is distinct from ``moiré-localized" states, which can be pinned by disorder or bound to individual moiré sites.
}
The work primarily focused on the 2$s$ exciton under fractional electron filling, without thoroughly addressing the 1$s$ and 3$s$ states. Yet the same theoretical approach can be readily extended to these higher or lower Rydberg excitons as well, motivating the more comprehensive analysis undertaken here.

{
Building on these observations, we additionally note that realistic moiré devices inevitably contain atomic-scale defects and are operated at finite temperatures. One key source of disorder in strongly correlated states is the presence of intrinsic atomic-scale defects. As established by extensive experimental studies, various point defects—such as chalcogen vacancies and topological defects—are commonly found in TMD monolayers, especially those grown by chemical vapor deposition (CVD)~\cite{Hong2015, Lin2015, Luo2024, Zhu2021}. Physically, these defects are known to act as localized charge traps by introducing mid-gap states within the band gap~\cite{Zhu2021, Chen2017, Yin2019}. This leads to the "pinning" of charge carriers in certain regions, breaking the perfect periodic arrangement of electrons in the moiré potential and seeding the formation of quasi-ordered domains~\cite{Kranjec2022, Mraz2023, Vodeb2024}.}
As a result, the excitonic resonances, especially those with large spatial extent, may undergo subtle shifts or broadening, reflecting the changed local environment within these quasi-ordered regions. Likewise, with rising temperature, thermal fluctuations grow in magnitude and can progressively undermine the long-range electronic order in moiré superlattices. In this regime, thermally activated carriers {move to random localized sites that are no longer part of the perfect electron crystal, weakening or entirely disrupting correlated insulating phases.} Such disordering not only diminishes the characteristic excitonic signatures but also induces redshifts or broadenings in exciton resonance peaks. Consequently, the pronounced moiré-induced excitonic features observed at low temperatures may gradually fade or disappear as the temperature increases~\cite{Xu2020,Fractional1_Regan,Burg2019,Wu2023,Matty2022,Ghiotto2021,Defect1_Koperski, Defect2_Cadiz}. These non-ideal effects complicate the interpretation of exciton spectra and can mask key correlation phenomena.

Motivated by these considerations, in this paper we significantly expand our previous theoretical framework to provide a more in-depth analysis of moir\'{e} excitons. We examine the Rydberg exciton series, particularly the 1$s$, 2$s$, and 3$s$ states~\cite{our_previous_work}.

\begin{figure}[t] 
    \centering
    \includegraphics[width=1.0\columnwidth]{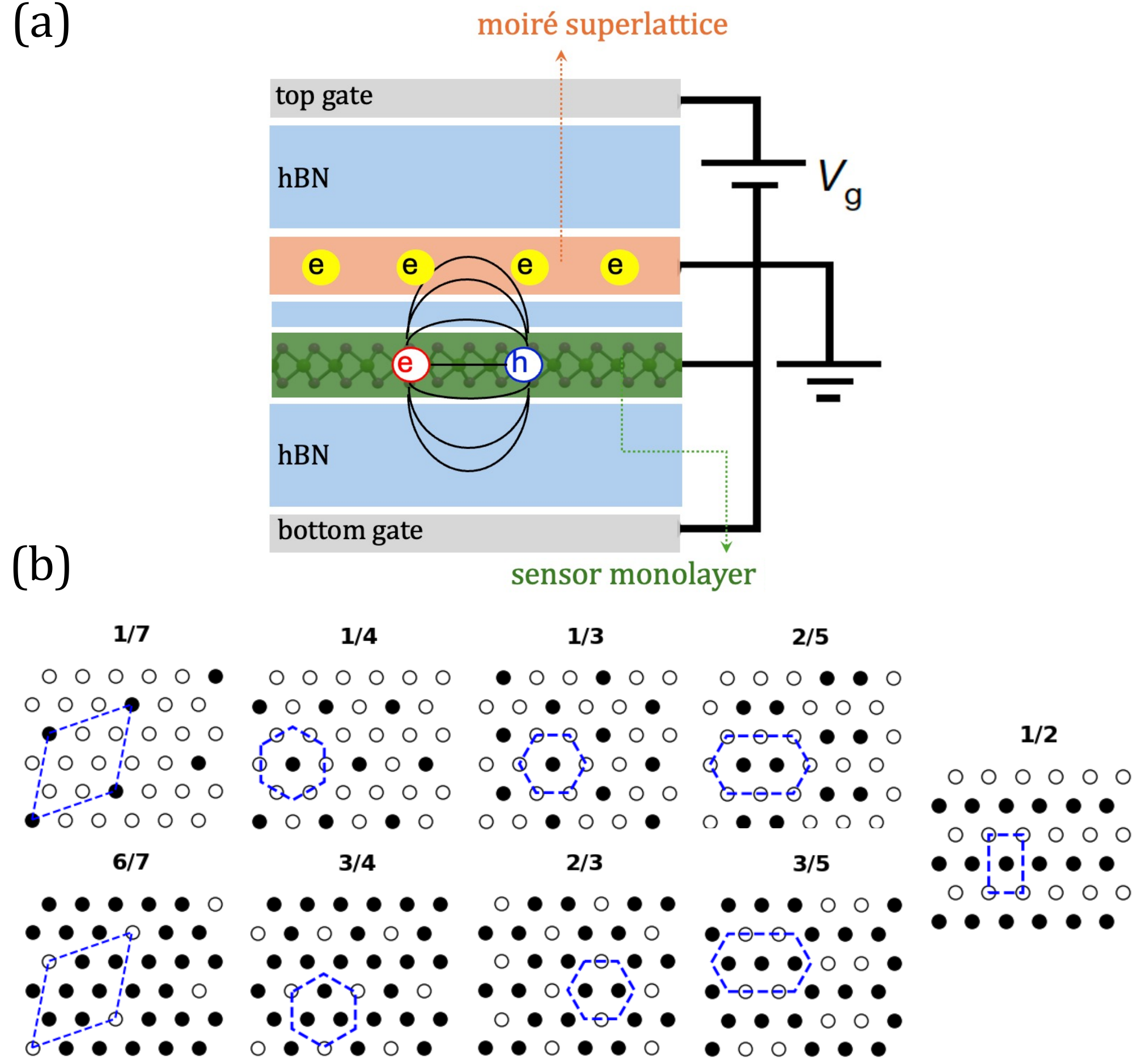}
    \caption{Schematic illustration of (a) the device structure and (b) representative electron configurations at various fractional fillings. The moir\'{e} superlattice (orange layer) is stacked above a separate “sensor” monolayer (green layer), and both layers are sandwiched between top and bottom hBN dielectrics. By tuning the gate voltages, one controls the fractional electron filling (\(\nu\)) in the moir\'{e} layer. For each fractional filling \(\nu\), white circles mark empty moir\'{e} lattice sites, and black dots indicate occupied sites. Blue dashed polygons highlight unit cells of each specific filling.}
    \label{fig:moire_schematic}
\end{figure}

We also investigate how fractional fillings and electron correlations in moir\'{e} superlattices influence the Rydberg exciton series. Specifically, we aim to clarify how different fractional fillings, including both idealized charge-ordered states and more realistic scenarios with quasi-disorder or finite temperatures, impact exciton states. Addressing this question is crucial because the relatively larger spatial extent of higher Rydberg excitons renders them particularly sensitive to changes in the moir\'{e} potential and local electron arrangements. At the same time, understanding such sensitivity is key to bridging idealized theoretical scenarios and experimental observations in devices where defects and thermal effects are unavoidable. Our analysis will thus illuminate not only the fundamental moir\'{e} exciton physics but also provide practical insights for designing future moir\'{e}-based optoelectronic systems. By treating various fractional fillings, and by systematically including perturbations from quasi-disorder and elevated temperatures, we lay the groundwork for a comprehensive understanding of how excitons can serve as probes of many-body correlation effects in TMDs heterostructure.

This paper is organized as follows. In Sec.~\ref{sec:Rydberg Excitons}, we study the fundamental differences among the 1$s$, 2$s$, and 3$s$ excitons in a moir\'{e} heterobilayer, highlighting how fractional electron filling affects each state. Section~\ref{sec:disorder and thermal} then introduces our approaches to incorporate defect-induced quasi-ordering and finite-temperature effects, presenting numerical results that compare ideal and disordered scenarios. We conclude in Sec.~\ref{sec:conclusion} by discussing the implications of our findings for both fundamental studies of strongly correlated excitons and their potential device applications. The Appendices include technical details
and parameter values used in the simulations.

\section{Rydberg Excitons under Fractionally Filled moir\'{e} Superlattices}
\label{sec:Rydberg Excitons}
In this section, we focus on how the emergent Wigner crystal states in moir\'{e} structure reshape the exciton as illustrated schematically in Fig.~\ref{fig:moire_schematic}. Our device structure in Fig.~\ref{fig:moire_schematic}(a) follows a dual-gated configuration similar to that of Refs.~\cite{Xu2020,moire_Zhang2022,BenMhenni2024}, where a WS\(_2\)/WSe\(_2\) moir\'{e} bilayer is closely coupled (via a thin hBN spacer) to a separate ``sensor'' WSe\(_2\) monolayer. By controlling the gate voltages, one can tune the electron filling in the moir\'{e} superlattice and detect it in the sensor WSe\(_2\). In essence, strong Coulomb interactions between the sensor excitons and the partially filled moir\'{e} layer enable optical readout of the fractional fillings.

We present a theoretical model that captures the interplay between the exciton {in the sensor monolayer} and ordered electrons in the moir\'{e} bilayer, illustrating how partial miniband filling and the resultant long-range Coulomb interactions can profoundly affect the formation and stability of excited exciton states. Specifically, we show that the ordered charge state induces characteristic modifications in the exciton wavefunctions, binding energies, and optical oscillator strengths.

\subsection{Model Hamiltonian}
\label{subsec:ModelHamiltonian}
We begin by describing an exciton whose center-of-mass coordinate is $\mathbf{R}$ and whose relative coordinate (between the electron and hole) is $\mathbf{r}$. The total Hamiltonian takes the form
\begin{align}
H \;=\; -\frac{\hbar^2 \nabla_{R}^{2}}{2M} \;-\; \frac{\hbar^2 \nabla_{r}^{2}}{2\mu}
\;+\; v(\mathbf{r})
\;+\;
V_{M}(\mathbf{R},\mathbf{r}),
\label{eq:H_total}
\end{align}
where \(M = m_e + m_h\) is the total mass of the electron--hole pair, \(\mu = \bigl(m_e^{-1} + m_h^{-1}\bigr)^{-1}\) is the exciton reduced mass, and \(v(\mathbf{r})\) is the intralayer electron--hole Coulomb interaction. The term
\( 
V_M(\mathbf{R},\mathbf{r})
\)
represents the external moir\'{e} potential generated by electrons that partially occupy the moir\'{e} lattice. We assume that $V_{M}(\mathbf{R} + i\mathbf{R}_1 + j\mathbf{R}_2,\mathbf{r}) = V_{M}(\mathbf{R},\mathbf{r})$ for integer $i,j$, reflecting the moir\'{e} superlattice periodicity defined by the translation vectors $\mathbf{R}_1$ and $\mathbf{R}_2$.

Bloch’s theorem tells us that, for a potential that is periodic in the center-of-mass coordinate $\mathbf{R}$, the exciton wavefunction can be written as
\begin{equation}
\Psi_{K}(\mathbf{R},\mathbf{r})
\;=\;
e^{i \mathbf{K}\cdot \mathbf{R}}\,
u_{K}(\mathbf{R},\mathbf{r}),
\label{eq:ExcitonBlochWavefunction}
\end{equation}
where $\mathbf{K}$ is the exciton crystal momentum, and $u_{K}(\mathbf{R},\mathbf{r})$ is a function periodic in $\mathbf{R}$. One may expand $u_{K}$ in reciprocal lattice vectors $\{\mathbf{G}\}$, $u_{K}(\mathbf{R},\mathbf{r})= \sum_{\mathbf{G}}\,u_{K}(\mathbf{G},\mathbf{r})\,e^{\,i\mathbf{G}\cdot \mathbf{R}}$.
Projecting the time-independent Schr\"odinger equation $H\Psi_{K} = E_{K}\Psi_{K}$ onto each plane-wave component $e^{\,i(\mathbf{K}+\mathbf{G})\cdot \mathbf{R}}$ yields a matrix eigenvalue problem in $\mathbf{G}$-space.

\begin{table}[t]
\centering
\setlength{\tabcolsep}{3pt}
\begin{threeparttable}
\caption{
Comparison of $E_{\alpha}$ for $\alpha = 1s,2s,3s$ excitons under various filling factors $\nu$. 
\textit{All Orbitals} employs the full set $\{1s,2s,2p_{\pm},3s,3p_{\pm}\}$, 
while \textit{Truncated Orbitals} restricts the calculation to the $s,p$ orbitals of same principal quantum number. 
All binding energies are in meV.
}

\label{tab:1s_2s_3s_table}
\begin{tabular}{c|ccc|ccc}
\toprule
\multicolumn{1}{c|}{\textbf{Filling }$\nu$} 
& \multicolumn{3}{c|}{\textbf{All Orbitals}} 
& \multicolumn{3}{c}{\textbf{Truncated Orbitals}}
\\
\cmidrule(lr){2-4}\cmidrule(lr){5-7}
& $E_{1s}$ & $E_{2s}$ & $E_{3s}$ 
& $E_{1s}$ & $E_{2s}$ & $E_{3s}$ 
\\
\midrule
1/7  & -130.07 & -13.45 & 12.03  & -130.00 & -13.04 & 11.96 \\
1/4  & -130.07 &  -8.97 & 19.53  & -130.00 &  -8.84 & 19.39 \\
1/3  & -130.07 &  -5.99 & 21.54  & -130.00 &  -5.96 & 21.53 \\
2/5  & -130.09 & -11.00 & 15.86  & -130.00 & -10.77 & 15.17 \\
1/2  & -130.10 &  -9.97 & 19.07  & -130.00 &  -9.82 & 18.76 \\
3/5  & -130.10 & -10.77 & 15.79  & -130.00 & -10.52 & 15.15 \\
2/3  & -130.07 &  -5.83 & 21.52  & -130.00 &  -5.78 & 21.52 \\
3/4  & -130.08 &  -8.81 & 19.55  & -130.00 &  -8.68 & 19.38 \\
6/7  & -130.08 & -13.53 & 12.04  & -130.00 & -13.15 & 12.00 \\
\bottomrule
\end{tabular}
\end{threeparttable}
\end{table}

A crucial aspect of moir\'{e}-modified excitons is that $V_{M}(\mathbf{R},\mathbf{r})$ can couple different internal orbital states (e.g., mixing $s$-type and $p$-type components) whenever the potential depends explicitly on the relative coordinate $\mathbf{r}$. To incorporate this effect, we expand $u_{K}(\mathbf{G},\mathbf{r})$ in a set of hydrogenic-like basis functions, $\varphi_\alpha(\mathbf{r})$ representing the $\alpha$ = 1$s$, 2$s$, 2$p_{\pm}$, 3$s$, 3$p_{\pm}$ states of an isolated 2D exciton. Thus, $u_{K}(\mathbf{G}, \mathbf{r}) \;=\; \sum_{\alpha} C_{\alpha}(\mathbf{G}) \,\varphi_{\alpha}(\mathbf{r})$. Substituting this expansion into the projected Schr\"odinger equation gives
\begin{align}
&\sum_{\mathbf{G}',\beta}
\Bigl\{
\Bigl[
\tfrac{\hbar^2\,(\mathbf{K}+\mathbf{G})^2}{2M}
\;+\;\varepsilon_{\beta}
\Bigr]\,
\delta_{\mathbf{G},\mathbf{G}'}
\,\delta_{\alpha,\beta} \nonumber \\ 
&\;+\;
V_{\alpha,\beta}(\mathbf{G}-\mathbf{G}')
\Bigr\}
\;C_{\beta}(\mathbf{G}')\;
=
E_{K}\;C_{\alpha}(\mathbf{G}),
\label{eq:ExcitonMatrix}
\end{align}
where $\varepsilon_{\beta}$ is the bare (hydrogenic) energy of orbital $\beta$, and $V_{\alpha,\beta}(\mathbf{G})$ encapsulates the coupling between different exciton orbitals. This mixing can produce significant shifts in the 2$s$ exciton resonance energy whenever the moir\'{e} lattice is fractionally filled with electrons, as we show in Table~\ref{tab:1s_2s_3s_table}. Further technical details of this derivation and parameters used are presented in Appendix~\ref{sec:Theoretical Details}.

We note that in principle one could include \emph{all} exciton orbitals (e.g., 1$s$, 2$s$, 2$p_{\pm}$, 3$s$, 3$p_{\pm}$) in a single diagonalization of Eq.~\eqref{eq:ExcitonMatrix}. However, to reduce computational cost we focus on the subset of orbitals relevant to a specific experimental resonance. For example, to analyze the 2$s$ exciton, we retain the 2$s$ and 2$p_{\pm}$ basis states, because those lie energetically close enough to hybridize strongly. Likewise, for the 3$s$ resonance, we might include the 3$s$ and 3$p_{\pm}$ states. As shown in Table~\ref{tab:1s_2s_3s_table}, our calculations show that including all orbitals in the same calculation does not qualitatively change the final 2$s$ (or 3$s$) resonance position.

\begin{figure}[t]
    \centering
    \includegraphics[width=1.0\columnwidth]{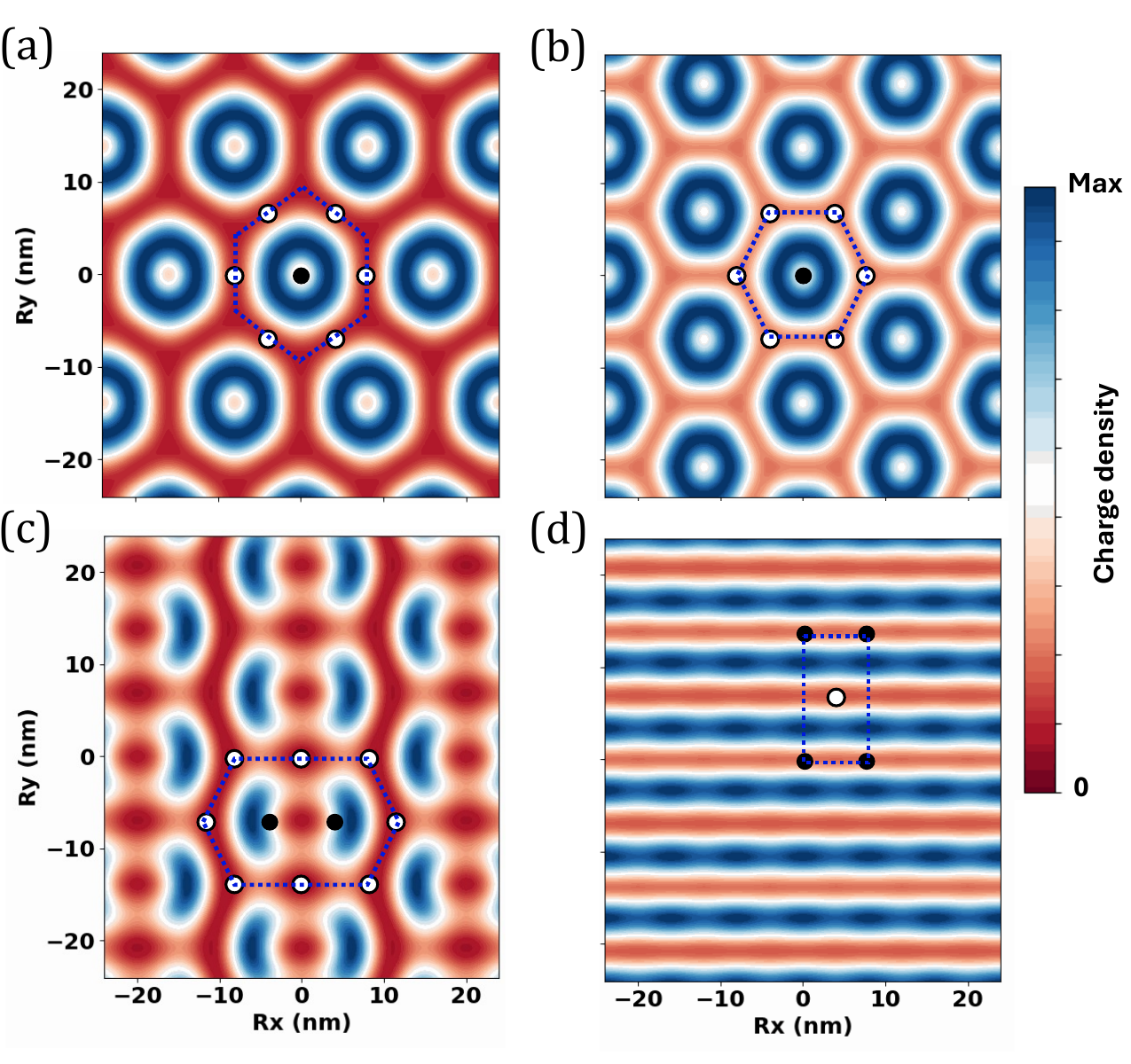}
    \caption{Center-of-mass probability distributions, $\rho(\mathbf{R})$, of the 2$s$ exciton at fractional fillings (a)~$\nu=1/4$, (b)~$\nu=1/3$, (c)~$\nu=2/5$, and (d)~$\nu=1/2$. The color scale indicates $\rho(\mathbf{R})$ from red (zero) to blue (maximum). These plots highlight how the 2$s$-like exciton wavefunction adapts to different ordered charge configurations in the moir\'{e} superlattice. {For each fractional filling $\nu$, white and black circles mark empty and occupied moiré lattice sites, respectively. Dashed-line polygons represent unit
cells of each specific filling.}}
    \label{fig:wavefunctions}
\end{figure}

\subsection{Probability distribution}
\label{sec:Probability distribution}
Once the coefficients $C_\alpha(\mathbf{G})$ have been determined by diagonalizing our Hamiltonian from Eq.(\ref{eq:ExcitonMatrix}), the center-of-mass probability distribution of the exciton can be obtained by integrating out the relative coordinate $\mathbf{r}$:

\begin{align}
\rho(\mathbf{R})
  &= \int d^{2}\mathbf{r}\,
     \bigl|\Psi_{\mathbf{K}=0}(\mathbf{R},\mathbf{r})\bigr|^{2} \nonumber\\[4pt]
  &= \sum_{\mathbf{G},\mathbf{G}',\alpha}
     C_{\alpha}(\mathbf{G})\,C_{\alpha}^{*}(\mathbf{G}')\,
     e^{i(\mathbf{G}-\mathbf{G}')\!\cdot\!\mathbf{R}}.
     \label{eq:COM_wavefunctions_definition}
\end{align}

Physically, $\rho(\mathbf{R})$ indicates where the exciton center-of-mass is likely to be found within one moir\'{e} supercell. For sufficiently strong moir\'{e} periods, these center-of-mass distributions can exhibit strong modulations that reflect how the fractional electron filling reshapes the Coulomb landscape.

\begin{figure*}[t] 
    \centering
    \includegraphics[width=2.0\columnwidth]{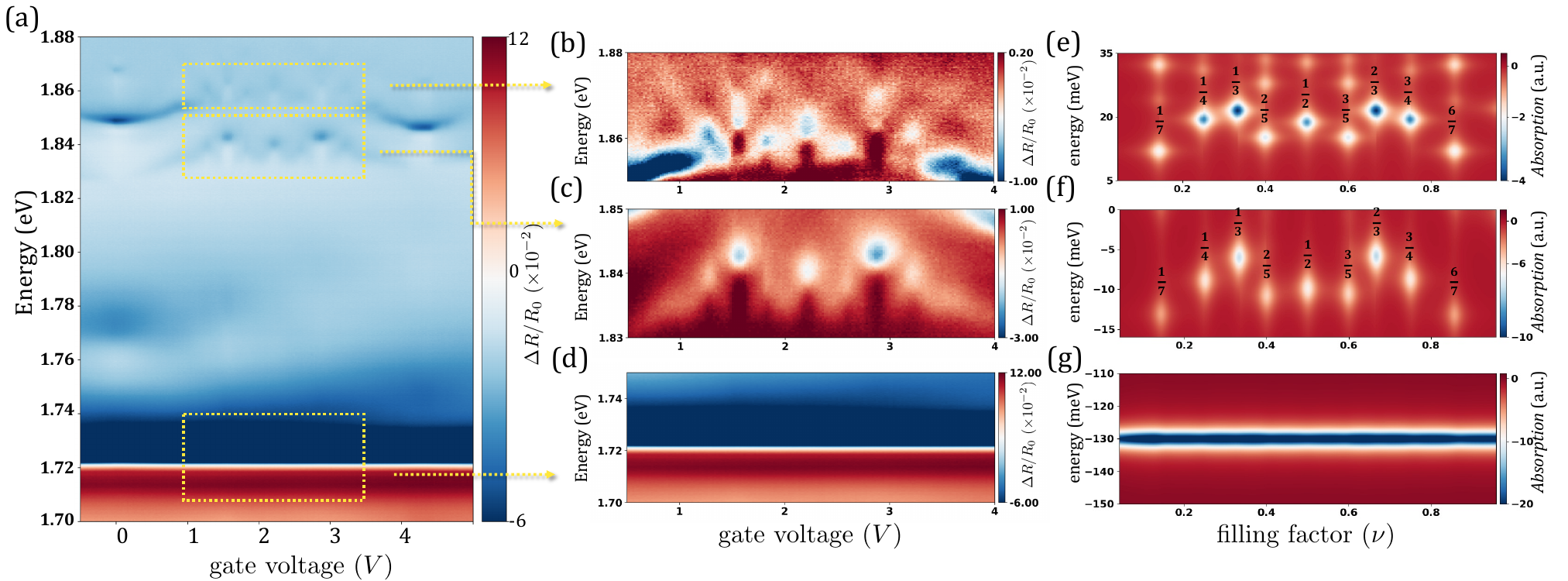}
    \caption{Comparison between experimental (from Ref.~\cite{Xu2020}) and theoretical exciton spectra.
    (a) Measured reflectance contrast ($R/R_{0}$) spectra over a broad energy range.  
    (b)--(d) Zoomed-in views of the same experimental data focusing on the 3$s$, 2$s$, and 1$s$ exciton energy ranges, respectively.  
    (e)--(g) Theoretically calculated absorption spectra (oscillator strength) for the 3$s$, 2$s$, and 1$s$ excitons, assuming an ordered electron arrangement in the moir\'{e} superlattice.  
    In each panel, the color scale is adjusted to distinguish the intensity levels clearly for each exciton state. Note that the 3$s$ and 2$s$ resonances undergo pronounced redshifts, whereas the 1$s$ exciton remains robust.
    }    
    \label{fig:moire_schematic2}
\end{figure*}

Figure~\ref{fig:wavefunctions} illustrates several representative examples of $\rho(\mathbf{R})$ for different fractional fillings, highlighting the interplay between the exciton in the sensor monolayer and ordered electrons. Red regions correspond to $\rho(\mathbf{R}) \approx 0$, whereas blue indicates a maximal value of $\rho(\mathbf{R})$. As shown, the exciton center-of-mass wavefunction can localize preferentially at specific sites or broaden into multiple overlapping shapes, depending on details such as inter-orbital mixing and the strength of the moir\'{e} potential.

\subsection{Comparison of experimental and theoretical results: 1$s$, 2$s$, and 3$s$ excitons}
As reported in Ref.~\cite{Xu2020}, reflection-contrast ($\Delta$$R$/$R_{0}$) measurements reveal distinct doping dependences of the 1$s$, 2$s$, and 3$s$ excitons. Figure~\ref{fig:moire_schematic2}(a) provides a broad overview of these experimental spectra over the relevant photon-energy range, while panels (b)--(d) zoom in on the 3$s$, 2$s$, and 1$s$ exciton spectral regions, respectively. 

In Fig.~\ref{fig:moire_schematic2}(d), the 1s exciton in the sensor WSe\(_2\) monolayer appears at lower energy (typically \(\sim\ 1.72\)\,eV in Refs.~\cite{MuellerMalic,Xu2020, Huang2016, stier2018}) and the energy position of the 1$s$ state is constant, and the reflection contrast  shows no significant modulation. This is evident as a uniform and flat feature across all gate voltages, indicating that the 1$s$ exciton remains largely unaffected by changes in electron filling. This doping insensitivity stems largely from the stronger Coulomb binding energy of the 1$s$ orbital, making a small point-like dipole less susceptible to the presence of electrons in the moir\'{e} structure as shown in Table.~\ref{tab:1s_2s_3s_table}.

In contrast, as shown in Fig.~\ref{fig:moire_schematic2}(c), the 2$s$ exciton exhibits a pronounced redshift and notable intensity modulation as the gate voltage is varied. Distinct dips and peaks appear at fractional fillings, corresponding to \(\nu = 1/4, 1/3, 1/2, 2/3, 3/4\), etc., highlighting the sensitivity of this Rydberg excitonic state. These resonances are influenced by the Coulomb potential from ordered charges in the moir\'{e} lattice, and thus, the response of 2$s$ exciton serves as a probe for the underlying correlated electron physics in the system. Similarly, the 3$s$ exciton response, shown in Fig.~\ref{fig:moire_schematic2}(b), reveals an energy redshift and stronger spectral modulations as a function of gate voltage. Physically, the larger spatial extent of higher Rydberg orbitals renders them more susceptible to variations in local electron density. Thus, whenever the moir\'{e} lattice is occupied at fractional fillings, the long-range Coulomb forces between excitons and charge carriers can induce notable modifications of the 2$s$ or 3$s$ exciton spectra, shifting their central energies by tens of meV and suppressing their amplitudes. While both 2$s$ and 3$s$ excitons exhibit doping-dependent shifts and quenching, the 3$s$ resonance is often weaker due to its weaker oscillator strength. Indeed, the 3$s$ binding energy is smaller, and the corresponding exciton radius is larger, so it experiences still stronger interaction with ordered charges on moir\'{e} superlattice. Overall, these observations confirm the hierarchy of doping sensitivity among excitonic states, with the 1$s$ exciton remaining stable while higher Rydberg states such as 2$s$ and 3$s$ respond strongly to fractional filling of the moiré superlattice.

\subsection{Absorption spectra}
\label{subsec:AbsorptionSpectra}

After solving the eigenvalue problem in Eq.~\eqref{eq:ExcitonMatrix}, we obtain a set of exciton eigenstates, where wavevector \(\mathbf{K}\) is a good quantum number. At low temperatures and in the linear optical regime, optically bright excitons typically reside near \(\mathbf{K} \approx 0\) (the light cone). Thus, we focus on these \(\mathbf{K} = 0\) states for the absorption spectrum. The absorption coefficient \(\alpha(\hbar\omega)\) is written as
\begin{eqnarray} 
\alpha(\hbar \omega) &\propto& \sum_j \frac{\left|\int_{} \Psi_{j}(\mathbf{R}, r=0) d^2\mathbf{R}\right|_{\mathbf{K} = 0}^2}{(\hbar \omega - E_{j})^2 + \delta^2} \nonumber \\
&\propto& \sum_j  \frac{ |\sum_{s}C_s(\mathbf{G}=0) \varphi_{s}(\mathbf{r}=0)|^2}{(\hbar \omega - E_{j})^2 + \delta^2},
\label{eq:Absorption}
\end{eqnarray}
where \(E_j\) is the eigenenergy of exciton state \(\Psi_j\), and \(\delta\) $= 2$~meV is a broadening parameter. Since \(\varphi_{p}(\mathbf{r}=0)=0\) for \(p\)-type orbitals, only the \(s\)-wave components of each hybridized state contribute to the linear absorption. As a result, partial mixing between the \(s\)- and nearby \(p\)-type orbitals can shift or diminish the effective resonance of the exciton in a filling-dependent manner. A detailed derivation and additional explanation of Eq.~\eqref{eq:Absorption} can be found in Appendix~\ref{sec:appendix absorption}.

Figures~\ref{fig:moire_schematic2}(e)--(g) present theoretically computed results illustrating how the \(3s\), \(2s\), and \(1s\) exciton resonances evolve with varying fractional filling \(\nu\). As confirmed by experiments, the \(1s\) resonance is essentially robust against doping, reflecting its strong binding energy and reduced spatial extent. In contrast, both the \(2s\) and \(3s\) resonances exhibit large doping-induced shifts, stemming from inter-orbital coupling to the moir\'{e} potential. The \(1s\) exciton absorption peak is typically an order of magnitude stronger than \(2s\) or \(3s\). This result aligns with the standard hydrogenic exciton model for 2D semiconductors, in which the 1$s$ state dominates the linear optical response due to its large wavefunction amplitude at $r = 0$. In particular, the strong electric dipole transition matrix element in this small-$\mathbf{r}$ regime endows the 1$s$ exciton with a significantly higher oscillator strength compared to higher Rydberg states. Although the oscillator strength of the 2$s$ resonance is comparatively weaker, it remains readily observable in reflectance, thus serving as a sensitive probe for fractional electron filling. 

Consistent with experiments~\cite{Xu2020} and our theoretical work, the strongest 2$s$ resonances appear at $\nu=1/3$ and $\nu=2/3$. Indeed, when $\nu$ is tuned to a fractional filling, the same spectral signatures seen in the 2$s$ resonance are observed (such as peak shifts and oscillator-strength suppression). We have also analyzed the \(3s\) exciton and the larger spatial extent of the 3$s$ exciton also comes with an intrinsically weaker oscillator strength relative to the 1$s$ and 2$s$ resonances. In experimental measurements, the 3$s$ peak can therefore be very faint compare to other 1$s$ and 2$s$, potentially complicating its identification. Furthermore, in certain doping regimes (e.g., near $\nu = 1/7$ or $\nu = 6/7$), the 3$s$ resonance can be overshadowed by the nearby 2$s$ excitonic feature in the reflectance spectra, making it even more difficult to resolve experimentally. A direct quantitative comparison using Table.~\ref{tab:1s_2s_3s_table} further underscores the sensitivity of the 3$s$ exciton. For instance, while the 2$s$ exciton redshift between fractional fillings (e.g., from $\nu=1/7$ to $\nu=1/3$) is on the order of 7--8~meV, the 3$s$ exciton can undergo a larger shift (about 9~meV) along with a pronounced reduction in peak intensity. Considering that the intrinsic 2$s$--2$p$ energy separation is about 10~meV and the 3$s$--3$p$ gap is 4~meV, these doping-induced shifts represent a significant energy scale for higher Rydberg states~\cite{zhu2023, stier2018}. Such pronounced behavior highlights why the 3$s$ exciton can serve as a particularly sensitive indicator of disorder and correlated electronic states in moir\'{e} superlattices.

These findings match experimental observations of distinctly different doping responses across the exciton series, and underscore that higher Rydberg excitons can be highly sensitive probes of electron correlations. In the next sections, we will apply the same absorption-coefficient approach introduced here to explore how defects and thermal fluctuations affect the Rydberg-exciton spectra of moir\'{e} superlattices.

{Before studying the roles of defects and temperature, we emphasize the unique behavior at $\nu=1/3$ and $2/3$ fillings. Our calculations reveal that triangular potential landscape  leads to a suppression of the 2s-2p hybridization. The direct evidence for this suppression is the optical response shown in Fig. 3(f). The oscillator strength, which is proportional to the s-orbital character of the wavefunction (as explained in Appendix B), is strongest at $\nu=1/3$ and $2/3$. This indicates that the exciton state retains a dominant s-character at this filling, meaning the s-p mixing is weaker compared to other fractional fillings.

From $\nu=1/7$ to 1/3, we notice the ``climb up" in energy which is a consequence of this sudden change in the hybridization mechanism. For other fillings (e.g., $\nu=1/4,2/5$), stronger hybridization pushes the energy of the s-like state downwards. At $\nu=1/3$, this downward push is suddenly weakened, causing the energy to lie at a relatively higher value and thus appear as an upward jump in the overall energy trend. The physics at very low fillings like $\nu=1/7$ is different again, being dominated by interactions with sparse, impurity-like charges rather than a well-formed crystal. It is the transition between these different physical regimes that gives rise to the complex energy dependence.
}

\begin{figure}[b] 
    \centering
    \includegraphics[width=1.0\columnwidth]{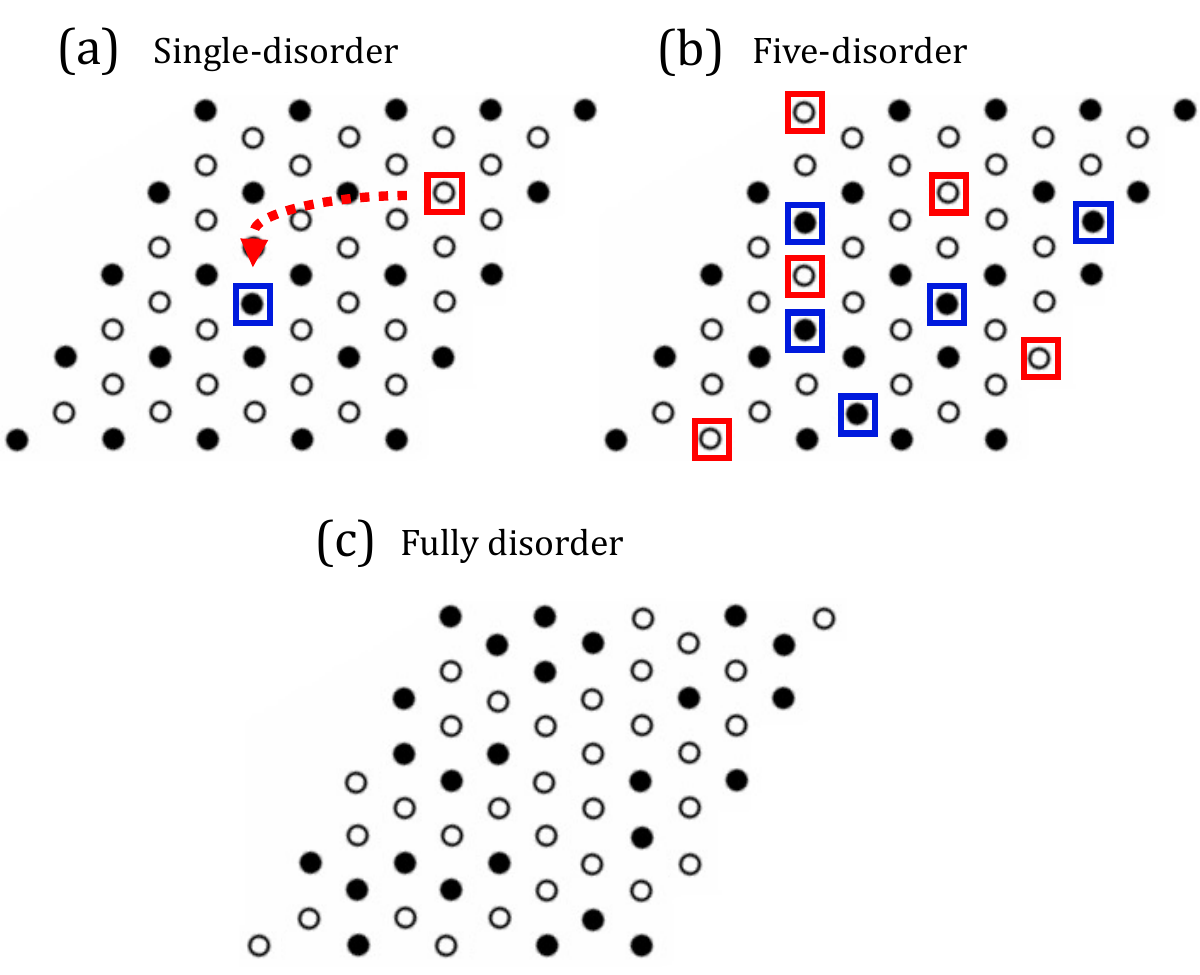}
    \caption{Schematic illustrations of different defect densities in a moir\'{e} lattice, highlighting how even a small number of displaced electrons can partially disrupt long-range order. For simplicity, a small section of the lattice is shown here, whereas our numerical calculations are performed on a \(12 \times 12\) moir\'{e} supercell based on the unit cell of each filling factor.
    (a)~Single-disorder: only one electron departs its original site (indicated by the red arrow and colored boxes), representing minimal quasi-ordering. (b)~Five-disorder: five electrons have left their original sites, illustrating a moderate level of quasi-ordering. (c)~Fully disordered: all electrons are randomly distributed without any long-range ordering. Black and white symbols represent filled and empty moir\'{e} site, respectively.
    }
    \label{fig:defect_schematic}
\end{figure}

\begin{figure*}[t] 
    \centering
    \includegraphics[width=2.0\columnwidth]{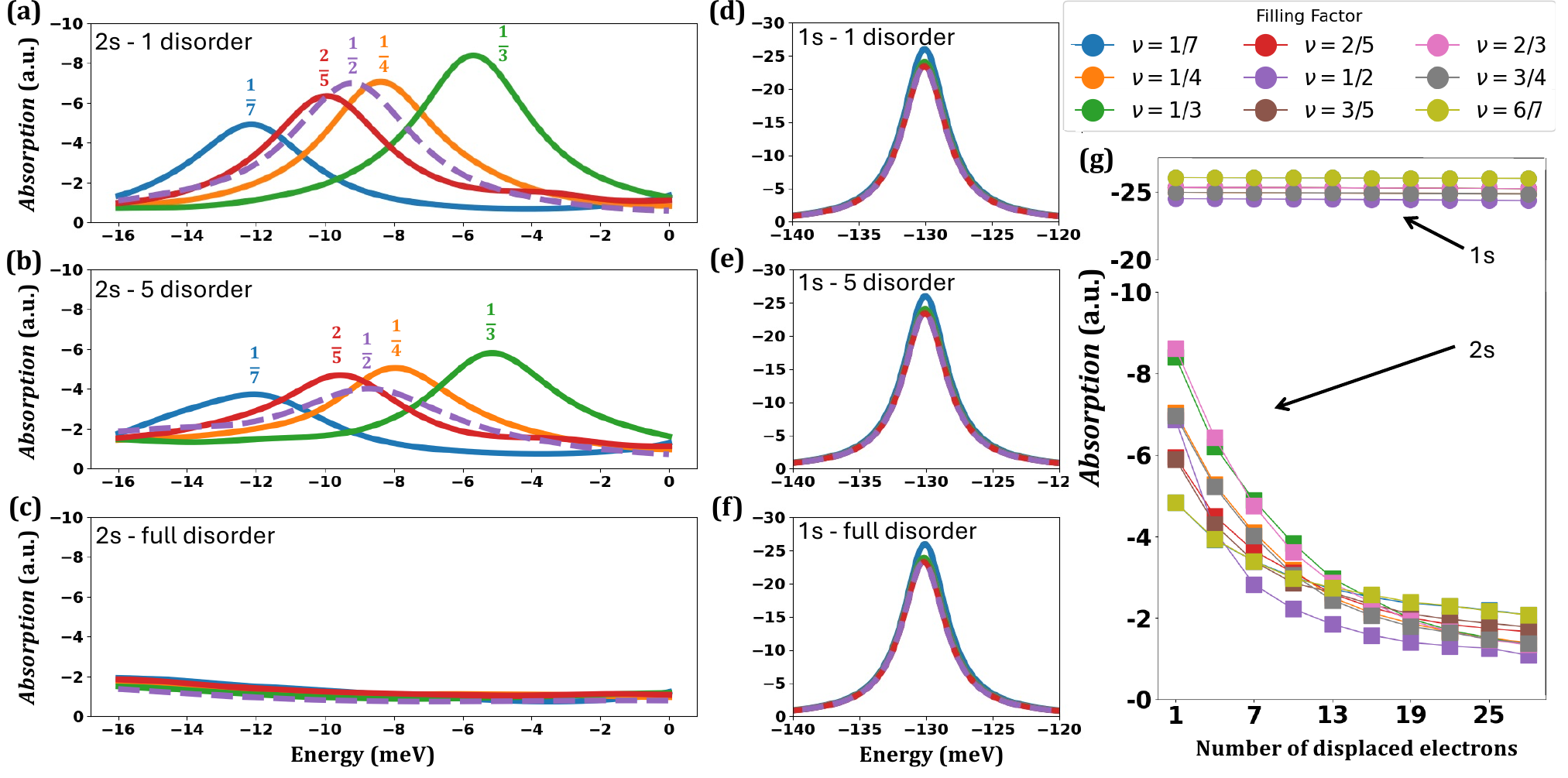}
    \caption{Impact of defect-induced quasi-ordering on exciton absorption in moir\'{e} superlattices. Simulated absorption spectra for the (a)--(c) 2$s$ and (d)--(f) 1$s$ excitons under different levels of structural disorder. Here, panels (a) and (d) show  single-disorder charge states; (b) and (e) correspond to a quasi-diorder state with five electrons displaced and (c) and (f) represent fully disordered electron distributions. Note that in panels (a)–(f), we show data only for filling factors $\nu$ from 1/7 to 1/2, as the corresponding spectra for $1-\nu$ are nearly identical to those for $\nu$.
    (g)~Summary of the absorption quenching due to defect-induced quasi-ordering, plotted for multiple fractional fillings $\nu$.  The horizontal axis indicates the total number of electrons displaced from their ideal moir\'{e} sites, and the vertical axis is the extracted absorption amplitude (a.u.).  Circles correspond to the 1$s$ exciton and squares to the 2$s$ exciton.  Different colors represent distinct filling factors from $\nu = 1/7$ to $\nu = 6/7$.  As the fraction of displaced electrons grows, higher-Rydberg excitons (2$s$) quickly lose their strong resonance features, whereas the 1$s$ exciton remains relatively unaffected.
}
    \label{fig:moire_schematic3}
\end{figure*}

\section{Defects and Thermal Effects}
\label{sec:disorder and thermal}
In realistic moir\'{e} TMD heterostructures, both structural defects and finite temperatures can significantly alter the long-range electronic order. We focus on two primary mechanisms: (i) \textit{defect-induced quasi-ordering}, where impurities or vacancies distort the ideal moir\'{e} potential and locally pin electron density, and (ii) \textit{thermal fluctuations} that can partially or completely melt charge-ordered states at higher temperatures. In the following sections, we will examine how the 1s and 2$s$ exciton resonances respond under these two scenarios, highlighting their distinct sensitivities to disorder and thermal effects.

\subsection{Defect-induced quasi-ordering}
\label{subsec:defect}
{

In this section, we model the effect of defect-induced charge inhomogeneity. As discussed in the introduction, atomic-scale defects in TMDs act as charge traps that can "pin" electrons, disrupting the ideal charge-ordered state. To simulate this physical pinning effect, we implement a quasi-ordering model where a controlled subset of electrons is displaced from its ideal, lowest-energy lattice sites. This local restructuring is particularly evident for 2\(s\) exciton, whose relatively large spatial extent makes it more sensitive to changes in the local electron configuration. Specifically, to} capture these effects in our theoretical framework, we construct large supercells that explicitly include defect sites. We then allow a subset of electrons to “leave” their ideal positions and relocate randomly elsewhere in the supercell as shown in Fig.~\ref{fig:defect_schematic}. By increasing the number of electrons that deviate from their original sites, we effectively model different defect densities and examine how the exciton resonances respond in each scenario. Details are explained in Appendix~\ref{sec:appendix_disorder}.

Our numerical calculations, illustrated in Fig.~\ref{fig:moire_schematic3}, show that quasi-ordering partially suppresses the exciton peaks but does so more moderately compared to the strong quenching observed under fully random doping. For example, Fig.~\ref{fig:moire_schematic3} compares three representative cases: 
\begin{enumerate}
    \item Only one electron has departed its original site, 
    \item Five electrons have departed, 
    \item A fully disordered distribution. 
\end{enumerate}
As the defect density grows, the exciton absorption systematically weakens for all considered filling factors. This trend indicates that the long-range potential pattern, which is responsible for the characteristic redshift and strong inter-orbital mixing, is increasingly disrupted. 

\begin{figure}[htbp!] 
    \centering
    \includegraphics[width=1.0\columnwidth]{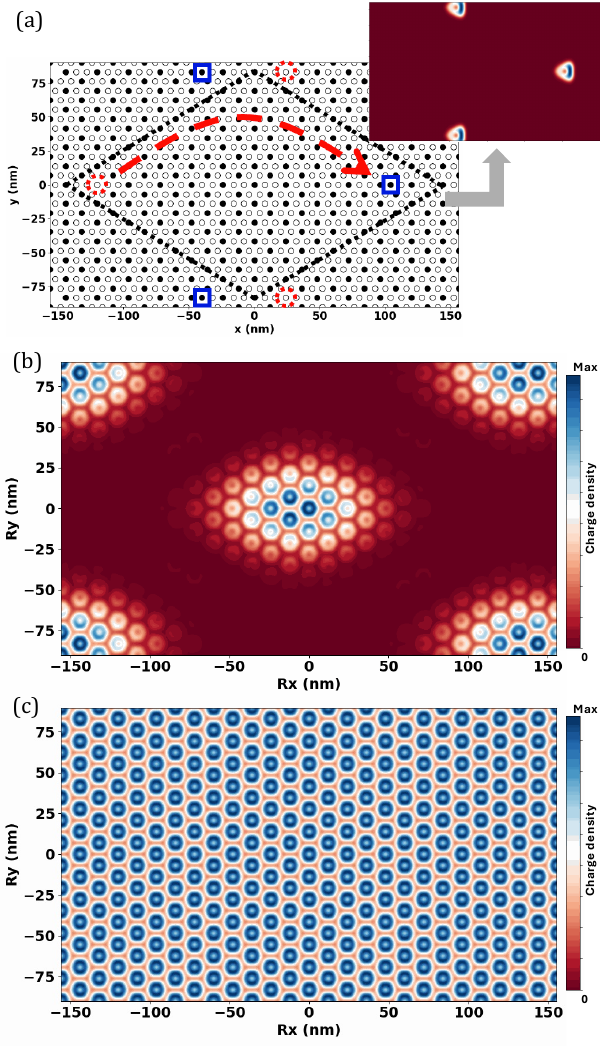}
    \caption{
(a) Schematic representation of a single disorder in a $12 \times 12$ moir\'{e} superlattice (black dashed line) at filling factor \( \nu = 1/3 \), where one electron departs from its original site, as indicated by the red arrow and colored boxes and circles. The inset shows the charge density of the lowest-energy state, which is localized near the blue boxes. The next-lowest state is localized near the red circle and has a similar energy to the lowest-energy state.
(b) Calculated Charge density using Eq.~(\ref{eq:COM_wavefunctions_definition}) of the 2$s$ state at \( \nu = 1/3 \) in the presence of the single disorder shown in (a), clearly showing wavefunction localization near the defect-free region. 
(c) Charge density of the 2$s$ state at \( \nu = 1/3 \) in a perfectly ordered system, showing uniform periodic features across the entire moir\'{e} pattern.
{The simulation domain shown here, which spans over 100 nm, is a realistic representation of the physical systems currently being studied. For instance, the seminal experimental works that our study is based on devices of similar scale~\cite{Xu2020, Fractional1_Regan}.}
}
    \label{fig:moire_defect_wavefunction}
\end{figure}

As illustrated in Fig.~\ref{fig:moire_defect_wavefunction}, introducing a single defect can disrupt the otherwise uniform electron arrangement at $\nu=1/3$ and partially localize the 2$s$ exciton wavefunction. In Fig.~\ref{fig:moire_defect_wavefunction}(b), we see that the exciton is pushed away from the defect region, whereas the defect-free system in Fig.~\ref{fig:moire_defect_wavefunction}(c) supports a regular, moir\'{e}-wide distribution of the 2$s$ orbital. This comparison confirms that even a minor imperfection can break the extended moir\'{e} potential pattern. Such effects become more pronounced when multiple defects are present or when temperature-driven fluctuations further disrupt the charge ordering.

Moreover, we observe that the presence of a single defect can result in nearly degenerate low-energy exciton states whose real-space wavefunctions localize in defect regions of the moir\'{e} superlattice. As indicated by the inset in Fig.~\ref{fig:moire_defect_wavefunction}(a), one such state is confined primarily to the site occupied by the displaced electron (blue box). Another state with similar energy (not shown) is localized at the vacated site (red circle). These states maintain orthogonality by shifting their amplitude distributions to avoid overlap, yet they share similar binding energies due to the comparable local potential landscapes.  Consequently, the localized exciton states exhibit closely spaced energy levels (-16.67 meV and -16.28 meV) and reduced oscillator strengths (Fig.~\ref{fig:moire_schematic3}) close to zero.

The energy of the state corresponding to Fig.~\ref{fig:moire_defect_wavefunction}(b) is --5.83 meV, and the absorption strength of the state is approximately 20 times larger than that of {the lowest-energy localized defect state (inset of Fig.~\ref{fig:moire_defect_wavefunction}(a))}, as shown in Fig.~\ref{fig:moire_schematic3}. Although the exciton state depicted in Fig.~\ref{fig:moire_defect_wavefunction}(b) lies at a higher energy, its larger spatial extent results in  a stronger contribution of the $\mathbf{G} = 0$ Fourier component than lower states, which leads to an enhanced absorption peak. In addition, the wavefunction distributions in Fig.~\ref{fig:moire_defect_wavefunction}(b) and (c) differ substantially; the corresponding energies and absorption intensities remain nearly unchanged, as can be seen in Fig.~\ref{fig:moire_schematic3}(a). Thus, the measured absorption profile remains robust. Consequently, these results indicate that small variations in defect distribution do not appreciably affect the experimentally observed absorption characteristics, underscoring the intrinsic stability of the system’s optical response. However, a significantly higher defect density in the moiré system can break this robust absorption peak. Moreover, the exciton wavefunction, which was extended under ideal conditions, becomes more localized in the presence of defects. Even relatively sparse defects can introduce localized pockets or domain boundaries across the moir\'{e} lattice, thereby partially pinning the electron distribution.

\begin{figure}[t] 
    \centering
    \includegraphics[width=1.0\columnwidth]{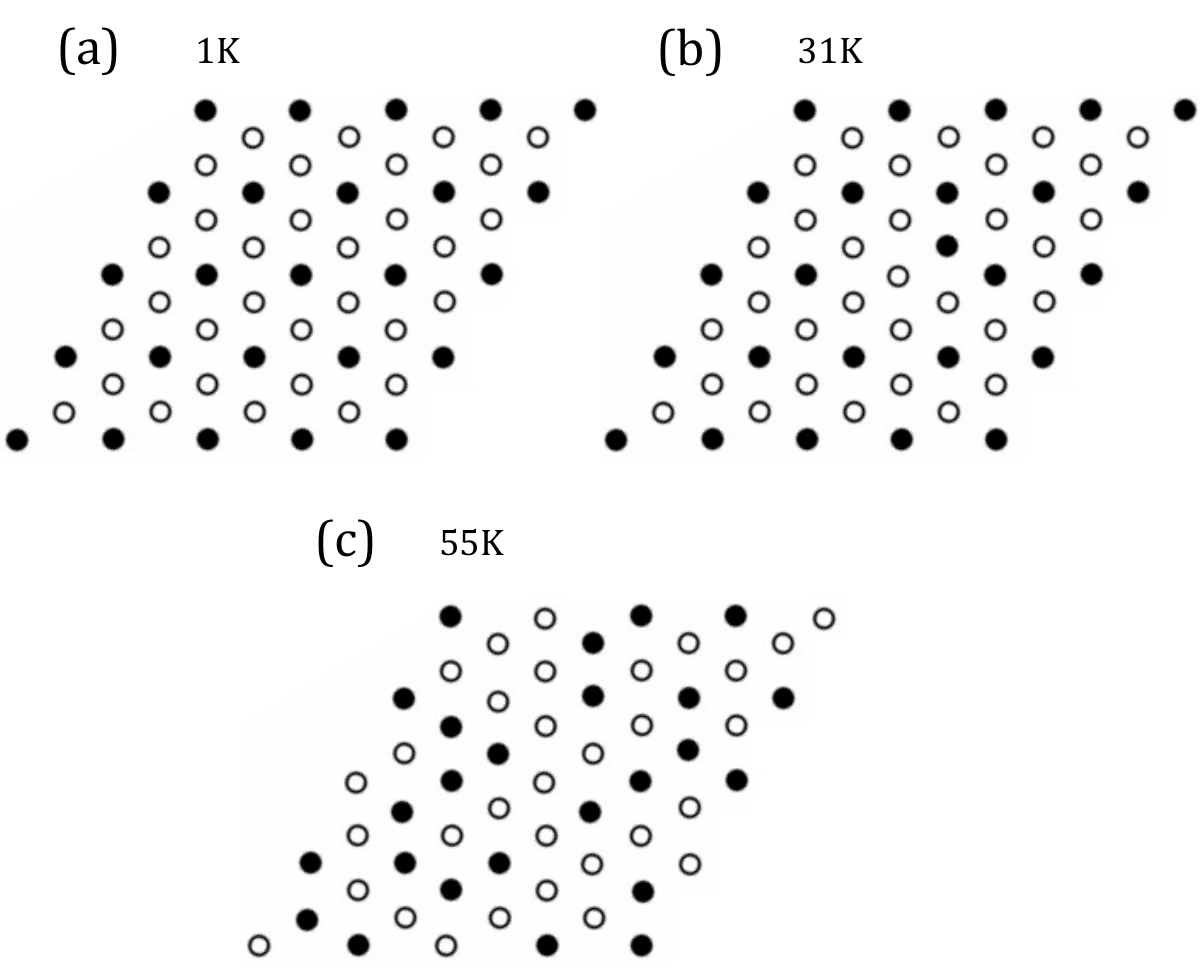}
    \caption{
Temperature-dependent evolution of charge ordering in  moir\'{e} lattice at a representative fractional filling of $\nu=1/3$, obtained via classical Monte Carlo simulations. 
(a)~At $T=1\,\mathrm{K}$, electrons occupy well-defined potential minima, forming a strongly correlated Wigner-like arrangement with minimal thermal agitation. 
(b)~At $T=31\,\mathrm{K}$, partial thermal activation allows electrons to move out of some minima, introducing noticeable disorder and reducing the spatial correlation length. 
(c)~By $T=55\,\mathrm{K}$, the increased thermal energy fully disrupts the long-range order, yielding a nearly random distribution of electrons. 
These snapshots illustrate the progressive melting of charge correlations as $k_{B}T$ becomes comparable to the moir\'{e} potential and electron--electron interaction scales.
    }
    \label{fig:thermal_schematic}
\end{figure}

\begin{figure*}[t] 
    \centering
    \includegraphics[width=2.0\columnwidth]{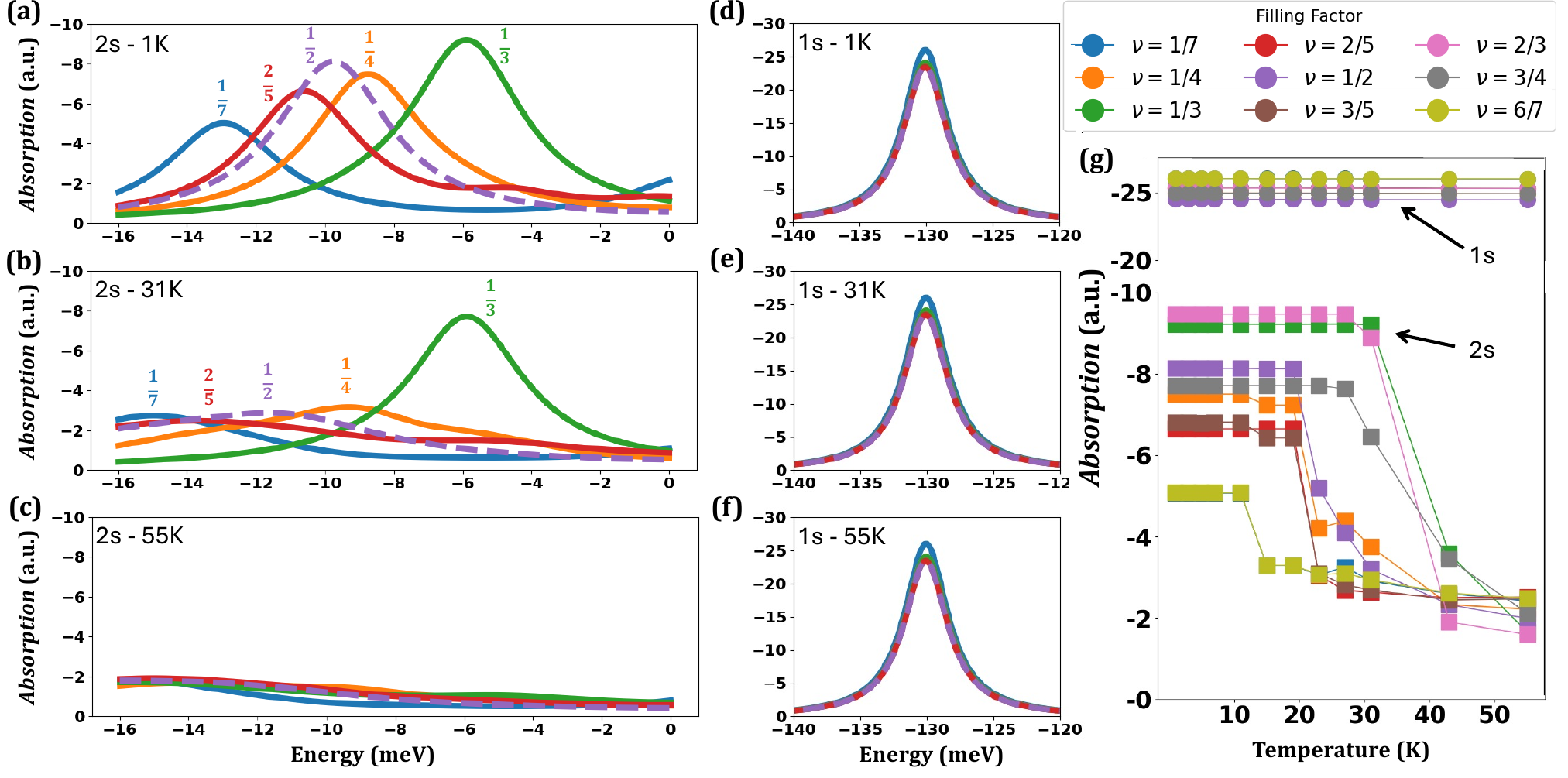}
    \caption{
    Calculated exciton absorption maps for different temperature $T$. 
(a)--(c) Theoretical absorption maps for the 2$s$ exciton at temperatures $T=1\,\mathrm{K}$, $31\,\mathrm{K}$, and $55\,\mathrm{K}$.  At low temperature (a), electrons occupy an ordered configuration in the moir\'{e} superlattice, yielding pronounced doping-dependent resonances.  With increasing $T$ (b) and (c), thermal fluctuations progressively destroy the long-range order and suppress the 2$s$ exciton features.  
(d)--(f) Analogous results for the 1$s$ exciton; its comparatively small spatial extent renders it more robust against thermal disorder, so its resonance varies less with $T$. Note that in panels (a)–(f), we show data only for filling factors $\nu$ from 1/7 to 1/2, as the corresponding spectra for $1-\nu$ are nearly identical to those for $\nu$.
(g)~Temperature-dependent evolution of the 1$s$ and 2$s$ absorption amplitudes for multiple filling factors $\nu$.  The horizontal axis indicates the lattice temperature, and the vertical axis is the simulated absorption peak (a.u.).  Different colors denote various fractional fillings, and circles (1$s$) or squares (2$s$) show the corresponding exciton resonance intensities.  Whereas the 2$s$ resonance rapidly weakens and eventually disappears as correlated electronic phases melt, the 1$s$ exciton remains comparatively stable up to higher temperatures.}
    \label{fig:moire_schematic4}
\end{figure*}

\subsection{Thermal fluctuations and melting of charge order}
\label{subsec:thermal}
In addition to structural imperfections, thermal energy can also drive the system toward disorder at sufficiently high temperatures. When the thermal energy scale ($k_{B}T$) becomes comparable to the electronic correlation or band dispersion scale (on the order of a few meV), long-range charge order in the moir\'{e} superlattice can be partially or completely destroyed \cite{Moire1_Tang, Bhowmik2024, Bistritzer2011, Huang2023, Zhou2024}. Even if the ground state at $T=0$ is a correlated insulator or Wigner-like crystal at fractional filling, modest increases in temperature (tens of Kelvin) can melt this ordered phase, leading to disordered electron distributions~\cite{Matty2022}.

Experimentally, reflection spectra collected over a temperature range from approximately 1.6\,K to 50\,K reveal that excitonic peaks, which signify strong exciton-ordered electron correlations, progressively weaken as temperature increases~\cite{Xu2020}. Notably, the correlated states at fractional fillings such as $\nu=1/3$ and $2/3$ can persist to relatively higher temperatures (around 26\,K) before melting, while other fillings tend to lose their ordered features at lower temperatures. This behavior suggests that different fractional fillings exhibit distinct critical temperatures, in line with the notion that each state possesses a characteristic correlation energy scale.

{
A critical question that arises is whether the observed quenching of the excitonic resonances stems from the thermal degradation of the exciton states themselves, rather than the melting of the charge order. This concern is addressed by comparing the relevant energy scales. The 2s exciton binding energy in WSe$_2$ is of the order of several tens of meV. In contrast, the highest temperature we consider where the correlated states melt is around $T=30 \sim 50 K$, which corresponds to a thermal energy of only $k_BT \approx 2.6 \sim 4.3$~meV. This thermal energy is considerably smaller than the energy required to either ionize the 2s exciton or excite the 2s exciton to a higher state. Therefore, the dominant finite-temperature effect captured in our work is not the thermal degradation of the exciton probe itself, but rather the melting of the generalized Wigner crystal state in the moiré layer. The dramatic suppression of the 2s absorption peak is a direct consequence of the loss of long-range charge order, which in turn averages out the periodic Coulomb potential ($V_M$) that causes the strong spectral features at low temperatures. This conclusion is further supported by the experimental data which show that the 2s exciton peak in the sensor monolayer remains clearly visible at 50 K, even as the moiré-induced features disappear~\cite{Xu2020}.
}

From a theoretical standpoint, thermal fluctuations smear out many of the excitonic spectral signatures that rely on a well-defined charge arrangement in the moir\'{e} structure. To model these effects, we adopt a classical Monte Carlo procedure (detailed in Appendix~\ref{sec:appendix_thermal}), wherein electron occupation on each moir\'{e} site is determined probabilistically based on Coulomb potential energies. As shown in Fig.~\ref{fig:thermal_schematic}, at 1 K, the Monte Carlo simulations indicate that electrons settle into strongly correlated moir\'{e}  states. As $T$ increases further, electrons become thermally activated out of potential minima, thus randomizing their spatial distribution. Consequently, by incorporating these temperature-dependent moir\'{e} fillings into our simulations, we find that the 2$s$ exciton resonance gradually diminishes and eventually disappears with increasing temperature, as shown in Fig.~\ref{fig:moire_schematic4}. Physically, the underlying mechanism is that once $k_{B}T$ exceeds the typical energy cost to move an electron between moir\'{e} sites, determined mainly by electron--electron interactions, long-range order can no longer be sustained. 

In this regime, exciton peaks associated with correlated phases undergo quenching or shifts at different rates, depending on each filling factor. Even without explicitly including fast carrier scattering by phonons, impurities, or electron--electron collisions, this randomization alone can account for substantial broadening and a reduction in peak intensity. Moreover, our temperature-dependent simulations indicate that each fractional filling factor undergoes a distinct melting temperature: excitonic peaks at certain fillings disappear earlier, whereas those at \(\nu=1/3\) and \(\nu=2/3\) persist until higher temperatures. As we discuss in Sec.~\ref{subsec:defect}, this sequential melting behavior contrasts with the more uniform degradation seen in defect-induced quasi-disorder. Such outcomes appear consistent with experimental results, where some fractional fillings lose their excitonic signatures before others, and overall the correlated resonances vanish by around 50\,K~\cite{Xu2020}.

{
Furthermore, it is crucial to justify our model for the high-temperature disordered phase. Instead of a uniform Fermi liquid, we model this state as a random distribution of localized point charges. This choice is physically motivated by the deep moiré potential, which is on the order of $\sim$100 meV~\cite{moireexciton_Wang2023}. Even at temperatures as high as 50 K ($k_BT \approx 4.3$~meV), the thermal energy is insufficient for most electrons to escape the confining potential wells and become fully delocalized. Therefore, even when the long-range Wigner crystal order melts, the electrons do not form a uniform Fermi liquid. Instead, the system is better described as a ``lattice gas", where electrons are still predominantly localized to individual moiré sites, but their arrangement across the lattice has become random.

The two different models for the disordered state predict qualitatively different spectral signatures. If we were to assume a uniform Fermi liquid, the delocalized charges would provide a uniform screening environment as the mobile electrons rearrange to screen the internal electric field of the excitonic dipole. This would subject the exciton to a spatially homogeneous electric field, resulting primarily in a Stark shift of its energy level. The absorption peak would therefore be expected to shift in energy, perhaps with some broadening, but would likely remain a discernible feature. In contrast, our model of randomly distributed but localized point charges creates a highly fluctuating, random potential landscape. The spatially extended 2s exciton wavefunction undergoes strong scattering from this rugged potential. This scattering mechanism is more effective at destroying the coherence of the excitonic resonance, leading to the severe quenching and broadening that causes the peak to be completely washed out, as seen in both our simulations Fig.~\ref{fig:moire_schematic3}(c), Fig.~\ref{fig:moire_schematic4}(c), and experimental observations at high temperatures~\cite{Xu2020}.
}

Interestingly, the 1$s$ exciton tends to remain more robust against these thermal fluctuations than 2$s$ states. Its smaller spatial extent (and thus stronger binding) makes it less sensitive to local variations in electron density, whereas the 2$s$ excitons, which have larger radii, display broader temperature-dependent shifts and greater susceptibility to thermally induced disorder. Consequently, understanding the temperature dependence of these excitonic features is critical for designing moir\'{e}-based optoelectronic devices that operate under realistic, finite-temperature conditions. 

\section{Conclusion and Outlook}
\label{sec:conclusion}
In this work, we have provided a comprehensive theoretical and numerical study of orbital-resolved excitonic states in fractionally-filled moir\'{e} superlattices made of transition-metal dichalcogenides. By systematically examining the 1$s$, 2$s$, and 3$s$ excitons under various fractional fillings, we have elucidated how correlated electron states markedly affect the higher Rydberg excitons, while the 1$s$ ground-state resonance remains relatively robust. Our analysis has shown that the 2$s$ and 3$s$ excitons undergo substantial energy shifts and changes in oscillator strength in response to small variations in electron filling, reflecting their larger spatial extent and enhanced sensitivity to moir\'{e} charge order.

We have also demonstrated how realistic device considerations, such as defect-induced quasi-ordering and finite-temperature effects, reduce or entirely suppress the excitonic signatures that arise from idealized charge-ordered phases. In particular, defects partially disorder the electron distribution, resulting in a progressive weakening of the redshift and oscillator-strength modifications. Moreover, thermal fluctuations can melt correlated insulating states at fractional fillings above certain critical temperatures. Our calculations show that this melting leads to the disappearance of the strong excitonic features observed at low temperatures, revealing distinct temperature scales for different fractional fillings.

Taken together, these results bridge the gap between ideal theoretical models and actual experimental observations, underscoring the interplay between exciton physics and strong correlated states in moir\'{e} heterostructures. Our findings shed light on a practical route to harness the sensitivity of higher Rydberg excitons as probes of emergent phases, while offering insight into how defects and finite temperatures alter exciton-based phenomena. 

An intriguing future direction is leveraging the tunable orbital structure of higher Rydberg excitons for terahertz (THz) optoelectronic applications. The energy spacing between the 2$s$ and 2$p$ excitonic states in monolayer TMDs can be on the order of a few meV (Appendix~\ref{sec:appendix_2s2p}), placing it squarely in the THz regime. By incorporating a dual-gated moir\'{e} device, one can systematically control the exciton binding energies and energy level spacing through vertical electric fields. This tunability provides a pathway for designing a voltage-controlled THz device. If one prepares a background population in the 2$s$ state (e.g., using a resonant pump), the 2$s$$\rightarrow$2$p$ transition can give rise to stimulated emission in the THz regime, offering an avenue for THz amplification. Conversely, a background occupation of 2$p$ excitons would enable the 2$p$$\rightarrow$2$s$ absorption process as a THz detection mechanism. Since the 2$p$ exciton is optically inactive in typical linear spectra, the transitions between 2$p$ and 2$s$ can be detected through optical resonance by incident THz light. Hence, one could envisage a moir\'{e}-based THz scale device, where the frequency is gate-tunable. We anticipate that the detailed microscopic understanding developed here will inform future studies of quantum many-body states in TMD-based moir\'{e} systems and inspire robust optoelectronic devices that exploit correlated excitonic effects for novel applications, including next-generation photonics and optoelectronics.

\begin{acknowledgments}
We are indebted to Jie Shan, Kin Fai Mak, and Yang Xu for sharing the experimental data and for fruitful discussions. This work is supported by the Department of Energy, Basic Energy Sciences, Division of Materials Sciences and Engineering under Award No. DE-SC0014349.
\end{acknowledgments}

\appendix
\section{Extended theoretical details}
\label{sec:Theoretical Details}
In this appendix, we provide a more in-depth derivation of the exciton Hamiltonian in reciprocal space, together with the procedure to incorporate orbital hybridizations under a moir\'{e} superlattice potential.

\subsection{Hamiltonian of exciton under moir\'{e} potential}
We begin with the total Hamiltonian of an exciton in a moir\'{e} superlattice, introduced in Eqs.~(\ref{eq:H_total})-(\ref{eq:ExcitonBlochWavefunction}) of the main text:

\begin{align}
H &= -\frac{\hbar^2 \nabla_{R}^2}{2M}
     -\frac{\hbar^2 \nabla_{r}^2}{2\mu}
     + v(\mathbf{r})
     + V_{M}(\mathbf{R},\mathbf{r}) \nonumber \\
  &= -\frac{\hbar^2 \nabla_{R}^2}{2M}
     + \hat{h}(\mathbf{r})
     + V_{M}(\mathbf{R},\mathbf{r}),
\label{eq:app_H_total}
\end{align}
where \(M = m_e + m_h = 0.65m_0\) is the total mass of the electron–hole pair where $m_0$ is the free-electron mass),~$\mu = \bigl(m_e^{-1} + m_h^{-1}\bigr)^{-1} = 0.16m_0$ is the exciton reduced mass~\cite{stier2018,RefS7}, and $v(\mathbf{r})$ is the intralayer electron–hole Coulomb interaction. Also, $\hat{h}(\mathbf{r}) \equiv -\frac{\hbar^2 \nabla_{r}^2}{2\mu}+v(\mathbf{r})$ and the operator represents the relative-motion Hamiltonian of the exciton.

We write a Bloch wavefunction
\begin{equation}
\Psi_{K}(\mathbf{R},\mathbf{r})
\;=\;
e^{\,i \mathbf{K}\cdot \mathbf{R}}\,
u_{K}(\mathbf{R},\mathbf{r}),
\label{eq:bloch wavefunction}
\end{equation}
with $u_{K}(\mathbf{R},\mathbf{r})$ being periodic in $\mathbf{R}$ on the moir\'{e} lattice. Expanding $u_{K}$ in reciprocal lattice vectors ${\mathbf{G}=\ell_1\mathbf{G_1}+\ell_2\mathbf{G_2}}$,
\begin{equation}
u_{K}(\mathbf{R},\mathbf{r})
\;=\;
\sum_{\mathbf{G}}\, u_{K}(\mathbf{G},\mathbf{r})
\, e^{\,i\mathbf{G}\cdot \mathbf{R}} ,
\end{equation}
and similarly expanding $V_{M}(\mathbf{R},\mathbf{r})$ in reciprocal space,
\begin{equation}
V_{M}(\mathbf{R},\mathbf{r})
\;=\;
\sum_{\mathbf{G}}\,V_{M}(\mathbf{G},\mathbf{r})
\, e^{\,i\mathbf{G}\cdot \mathbf{R}},
\end{equation}
one can project the Schr\"odinger equation onto each plane-wave component $e^{\,i(\mathbf{K}+\mathbf{G})\cdot \mathbf{R}}$ to obtain a matrix eigenvalue problem in $\mathbf{G}$-space. Denoting the exciton eigenenergy by $E_K$, we arrive at:
\begin{align}
\left[
\frac{\hbar^2(\mathbf{K}+\mathbf{G})^2}{2M}
\;+\;
\hat{h}(\mathbf{r})
\right]
\,u_{K}(\mathbf{G},\mathbf{r})
\;+\;\nonumber \\
\sum_{\mathbf{G}'}
V_{M}(\mathbf{G}-\mathbf{G}',\mathbf{r})
\,u_{K}(\mathbf{G}',\mathbf{r}) 
=E_{K}\,u_{K}(\mathbf{G},\mathbf{r}),
\label{eq:HGGequation}
\end{align}

which can be diagonalized numerically, keeping a sufficient number of reciprocal vectors to converge. The number of reciprocal lattice vectors ($ \mathbf{G} = \ell_1 \mathbf{G_1}+ \ell_2 \mathbf{G_2}$) in the simulations is defined by all integers in the range $-N_G \leq \ell_1, \ell_2 \leq N_G$. We find that $N_G = 10$ is sufficient to achieve converged results in terms of energy eigenvalues.\\

\subsection{Exciton-orbital hybridization}
If the moir\'{e} potential $V_{M}(\mathbf{R},\mathbf{r})$ depends on the relative coordinate $\mathbf{r}$, it can couple different orbital states of the exciton. For example, a $2s$ exciton can mix with $2p$ orbitals if the potential has the appropriate spatial variation. In general, we expand $u_{K}(\mathbf{G},\mathbf{r})$ in a set of hydrogenic-like basis functions $|\,\alpha\,\rangle \equiv \varphi_{\alpha}(\mathbf{r})$:
\begin{equation}
u_{K}(\mathbf{G},\mathbf{r})
\;=\;
\sum_{\alpha}\, C_{\alpha}(\mathbf{G}) \,\varphi_{\alpha}(\mathbf{r})
~\equiv~ \sum_{\alpha} C_{\alpha}(\mathbf{G}) \,\vert \alpha \rangle,
\label{eq:hydrogen basis function}
\end{equation}
where \(\varphi_{\alpha}(\mathbf{r})\) represents a 2D hydrogen-like eigenstate labeled by the principal and angular-momentum quantum numbers (\(n, \ell\)). In practice, we include the relevant set of orbitals needed to capture key resonances.
This transforms the exciton Hamiltonian:
\begin{align}
\sum_{G'} \biggl\{
\Bigl(\tfrac{\hbar^2 (\mathbf{K} + \mathbf{G})^2}{2M} \;+\; \hat{h}(\mathbf{r})\Bigr)\,\delta_{\mathbf{G},\mathbf{G'}} 
\;+\; V(\mathbf{G}-\mathbf{G'},\,\mathbf{r})
\biggr\}\nonumber \\
\times\sum_{\beta} C_{\beta}(\mathbf{G'})\,\lvert \beta\rangle
\;=\;
E \;\sum_{\gamma} C_{\gamma}(\mathbf{G})\,\lvert \gamma\rangle,
\end{align}

\begin{align}
\sum_{\mathbf{G}',\beta}
\bigg[\bigg[
\frac{\hbar^2(\mathbf{K}+\mathbf{G})^2}{2M} + \varepsilon_{\beta}
\bigg]\delta_{\mathbf{G},\mathbf{G}'}\delta_{\alpha,\beta}
\;+\;
V_{\alpha,\beta}(\mathbf{G}-\mathbf{G}')
\bigg]\,\nonumber \\
\times C_{\beta}(\mathbf{G}')
\;=\; 
E\,C_{\alpha}(\mathbf{G}),
\end{align}
where $\varepsilon_{\beta}$ is the bare hydrogenic energy of the $\beta$th orbital.
Based on experimental data, we define the following energy differences for the excitonic states:
$E_{1s}-E_{2s}=-130~\mathrm{meV},E_{2s} - E_{3s} = -22~\mathrm{meV},E_{2s} - E_{2p} = 10~\mathrm{meV},E_{3s} - E_{3p} = 4~\mathrm{meV}$~\cite{zhu2023,stier2018}.

Next, to calculate $V_{\alpha,\beta}(\mathbf{G}-\mathbf{G}')$ let's consider there are electrons at positions \(\mathbf{R}_{M}\) in a layer separated from the TMD monolayer by a distance \(d\). The real-space potential for the interaction between moir\'{e} electrons and an exciton has the form
\begin{align}
V_{M}(\mathbf{R}, \mathbf{r})
&= \sum_{\mathbf{R}_{M}} \frac{e^{2}}{\varepsilon}\,
\Biggl[\,
\frac{1}{\sqrt{\,d^{2} 
   + \bigl(\mathbf{R} + \tfrac{1}{2}\,\mathbf{r} - \mathbf{R}_{M}\bigr)^{2}}}
\nonumber\\[5pt]
&\quad\;\;-\;
\frac{1}{\sqrt{\,d^{2} 
   + \bigl(\mathbf{R} - \tfrac{1}{2}\,\mathbf{r} - \mathbf{R}_{M}\bigr)^{2}}}
\Biggr],
\label{eq:V_R_r_moire}
\end{align}
where we adopt $\epsilon = 5$ as the static dielectric constant of hBN, and $e$ is the elementary charge. The distance (d) between moir\'{e} layer and exciton plane is 4~nm. Then, the reciprocal-space representation of the potential is obtained by Fourier transform,
\[
V_{M}(G, \mathbf{r})
\;=\;
\frac{1}{A} \,\int V_{M}(\mathbf{R}, \mathbf{r})\,e^{-\,i\,\mathbf{G}\cdot \mathbf{R}}\,d\mathbf{R},
\]
where \(A\) is the total Moir\'{e} area. A more explicit calculation yields
\begin{align}
V_{M}(\mathbf{G}, \mathbf{r}) \;=\; -\,\frac{2\pi\,e^2}{\varepsilon \,A}\;\frac{e^{-\,d\,G}}{G}\;
\Bigl(\,e^{-\tfrac{i}{2}\,\mathbf{G}\cdot \mathbf{r}} \;-\; e^{\tfrac{i}{2}\,\mathbf{G}\cdot \mathbf{r}}\Bigr)\nonumber \\
\times\;\biggl(\sum_{\mathbf{R}_M}\,e^{-\,i\,\mathbf{G}\cdot \mathbf{R}_M}\biggr),
\label{eq:V_of_G_r_result}
\end{align}
In the above, \(e^{\mp \tfrac{i}{2}\,\mathbf{G}\cdot \mathbf{r}}\) shows that electron and hole in the exciton 
carry opposite signs under the external charge; the factor \(e^{-\,d\,G}\!/G\) comes from integrating over 
\(\sqrt{\,d^2 + R^2}\) in the plane, and 
\(\sum_{\mathbf{R}_M} e^{-\,i\,\mathbf{G}\cdot \mathbf{R}_M}\) encapsulates the form factor of the 
 moir\'{e} electron distribution. In principle, a rigorous treatment of the exciton would account for the distinct positions of the electron and hole, yielding an interaction term of the form,
$\Bigl(e^{\tfrac{i}{2}\,\mathbf{G}\cdot \mathbf{r_e}} 
      \;-\; 
      e^{\tfrac{i}{2}\,\mathbf{G}\cdot \mathbf{r_h}}\Bigr)$,
where \(\mathbf{r}_e\) and \(\mathbf{r}_h\) are the electron and hole coordinates, respectively.
However, since the electron and hole masses are not vastly different, 
it is often sufficient to assume $\mathbf{r}_e \approx \frac{r}{2}, \mathbf{r}_h\approx -\frac{r}{2}$ and thereby 
treat the exciton as a single entity with opposite charge contributions. This is a sound approximation for most practical conditions, 
and deviations in the final results due to ignoring the exact 
electron–hole separation are insignificant for the purposes of this work.

Finally, by projecting \(V(G,\mathbf{r})\) onto the hydrogenic orbitals, we get the matrix elements of the moir\'{e} potential between exciton orbitals $\alpha$ and $\beta$:
\begin{equation}
V_{\alpha,\beta}(\mathbf{G}-\mathbf{G}')
\;=\;
\langle \alpha|\,V_{M}(\mathbf{G}-\mathbf{G}',\mathbf{r})\,|\beta\rangle.
\end{equation}

\subsection{Hydrogen-like exciton states}
\label{sec:hydrogen_like}
Our calculations employ a simplified hydrogenic (2D hydrogen-like) model for the orbital wavefunctions of the excitons, including the ground state (1\(s\)) and higher Rydberg states (\(2s\), \(2p_{\pm}\), \(3s\), \(3p_{\pm}\)). This choice is motivated by the large exciton binding energies in TMD monolayers and the well-established success of hydrogen-like approaches in describing their radial wavefunctions~\cite{MuellerMalic}.

Each 2D hydrogen-like orbital can be written as a product of a radial part and an angular phase factor. For instance, the \(s\)-wave states have no angular dependence, while \(p_{\pm}\) states carry \(e^{\pm i\theta}\). Below are 2D hydrogen-like orbital, \(\phi_{1s}, \phi_{2s}, \phi_{2p_{\pm}}, \phi_{3s}, \phi_{3p_{\pm}}\)~\cite{Yang_2DH_1991}:
\begin{align*}
\phi_{1s}(r,\theta)
&=\;
\frac{\beta_{1}}{\sqrt{2\pi}}\,
e^{-\tfrac{\beta_{1}r}{2}}, 
\\[6pt]
\phi_{2s}(r,\theta)
&=\;
\frac{\beta_{2}}{\sqrt{3}}
\Bigl(1 - \beta_{2}r\Bigr)\,e^{-\tfrac{\beta_{2}r}{2}}
\,\frac{1}{\sqrt{2\pi}},
\\[6pt]
\phi_{2p_{\pm}}(r,\theta)
&=\;
\frac{\beta_{2}^{2}}{\sqrt{6}}\,r\,
e^{-\tfrac{\beta_{2}r}{2}}
\,\frac{1}{\sqrt{2\pi}}\,
e^{\pm i\theta},
\\[6pt]
\phi_{3s}(r,\theta)
&=\;
\frac{\beta_{3}}{2\sqrt{5}}
\Bigl(2 - 4\beta_{3}r + \beta_{3}^{2}r^{2}\Bigr)
e^{-\tfrac{\beta_{3}r}{2}}
\,\frac{1}{\sqrt{2\pi}},
\\[6pt]
\phi_{3p_{\pm}}(r,\theta)
&=\;
\frac{\beta_{3}^{2}}{\sqrt{30}}\,r\,
\bigl(3 - \beta_{3}r\bigr)\,
e^{-\tfrac{\beta_{3}r}{2}}
\,\frac{1}{\sqrt{2\pi}}
\,e^{\pm i\theta}.
\end{align*}
The constants \(\beta_{n}\) set the length scale of the \(n\)th orbital and are chosen to reproduce known exciton binding energies or average radii in TMD monolayers. In this work, we select \(\beta_n\) values to satisfy \(\langle r_{1s}^2\rangle = 1.7\,\mathrm{nm}\), \(\langle r_{2s}^2\rangle = 6.6\,\mathrm{nm}\), and \(\langle r_{3s}^2\rangle = 14.3\,\mathrm{nm}\)~\cite{stier2018}, through the relation $\langle r_{ns}^2\rangle=\int d^{2}{r}\,~{r}^2\bigl|\phi_{ns}({r})\bigr|^{2}$. 

The external potential \(V(\mathbf{R}, \mathbf{r})\) couples different orbitals (\(\alpha \leftrightarrow \beta\)) when its Fourier components transfer momentum \(\mathbf{G}-\mathbf{G}'\). In particular, \(s\)--\(p\) hybridization can substantially modify the exciton spectrum. Numerically, one retains a sufficient set of reciprocal vectors to achieve convergence.

By accounting for multiple hydrogen-like orbitals, one captures the formation of moir\'{e} exciton “minibands” and the interplay between orbitals, particularly creating new states by mixing 2$s$ and 2$p$. This richer orbital resolution is crucial for describing how fractional electron fillings renormalize each exciton resonance differently. As shown in the main text, higher Rydberg states (2\(s\), 3\(s\)) are often more sensitive to ordered charge in moir\'{e} superlattices and therefore can exhibit pronounced shifts or quenching.\\

\section{Absorption spectrum}
\label{sec:appendix absorption}
In a moir\'{e} superlattice, the exciton eigenstates (labeled by filling factor $\nu$ and wavevector $\mathbf{K}$) can generally be written as
\begin{equation}
    \Psi^\nu_{\mathbf{K}}(\mathbf{R},\mathbf{r})
    \;=\;
    \sum_{G,\alpha}\,
        e^{\,i\,(\mathbf{K}+ \mathbf{G})\cdot \mathbf{R}}\,
        C_\alpha^\nu(\mathbf{G})\,
        \varphi_\alpha(\mathbf{r})\,,
    \label{eq:psi_nu}
\end{equation}
where
\begin{itemize}
    \item $\mathbf{R}$ is the exciton center-of-mass coordinate,
    \item $\mathbf{r}$ is the relative coordinate of electron--hole,
    \item $\mathbf{G}$ are reciprocal-lattice vectors of the moir\'{e} superlattice,
    \item $C_\alpha^\nu(\mathbf{G})$ are coefficients determined by solving the eigenvalue problem for the exciton in a periodic potential,
    \item $\varphi_\alpha(r)$ are hydrogen-like basis orbitals.
\end{itemize}

We now specialize to the case of optical absorption ($\mathbf{K} \approx 0$) and focus on the probability amplitude of the exciton wavefunction at $\mathbf{r}=0$, denoting the overlap between electon and hole of exciton.  Concretely, the oscillator strength of the exciton state $\Psi^\nu$ can be written as
\begin{align}
   \Bigl|\int
       \Psi^\nu\bigl(\mathbf{R},r=0\bigr)d^2\mathbf{R}\;
   \Bigr|^2_{\mathbf K=0} .
   \label{eq:osc_strength_definition}
\end{align}
Substituting the general expansion of Eq.~\eqref{eq:psi_nu} at $r=0$ yields
\begin{align}
   \biggl|\int d^2 \mathbf{R}
   \,
     \sum_{\mathbf{G},\alpha}
       e^{\,i\,\mathbf{G}\cdot \mathbf{R}}
       \,
       C_\alpha^\nu(G)\,
       \varphi_\alpha(\mathbf{r}=0)
   \biggr|^2.
   \label{eq:osc_str_interm}
\end{align}
Because $\varphi_{p\pm}(r=0) = 0$ (and similarly for $d$ states, \emph{etc.}), the \emph{only} term that contributes in the sum over $\alpha$ is an $s$-type orbital.  Let us denote $\alpha=s$ for simplicity.  Moreover, the integral $\int d^2 \mathbf{R}\, e^{\,i(G - G')\cdot \mathbf{R}}$ is non-zero only if $G=G'$, and in particular the $G=0$ channel plays a key role in absorbing zero in-plane momentum photons.  Hence we arrive at the simplified expression
\begin{align}
Oscillator~Strength &\propto
   \bigl|
   \sum_{s}
     C_s^\nu(\mathbf{G}=0) \varphi_{s}(\mathbf{r}=0)
   \bigr|^2,
   \label{eq:f_G0_final}
\end{align}
The key result is that the oscillator strength of an $s$-like exciton state is set by the zero-momentum ($\mathbf{G}=0$) component of its wavefunction expansion. Equation~\eqref{eq:f_G0_final} shows why doping can drastically reduce the oscillator strength of the $2s$ resonance:  once the $s$ state mixes with nearby $p$ states (which do not contribute at $r=0$), the $s$-type amplitude at $\mathbf{G}=0$ shrinks.  This lowered $2s$-component directly corresponds to a suppressed absorption peak.  If the moir\'{e} electrons form an especially fractional filled configuration, the potential $V_{M}$ can enhance certain $s$--$p$ couplings and shift or reshape the exciton resonance accordingly. By incorporating Eq.~\eqref{eq:f_G0_final} into the Lorentzian form of the absorption coefficient in Eq.~\eqref{eq:Absorption}, we obtain a description of how the exciton absorption spectrum depends on the filling factor $\nu$.

\section{Disorder simulation}
\label{sec:appendix_disorder}

In the main text (Sec.~\ref{subsec:defect} and \ref{subsec:thermal}), we discussed how real devices often contain structural imperfections or finite-temperature effects that can partially disrupt the ideal moir\'{e}-ordered electronic configuration. In this Appendix, we provide additional technical details on how we incorporate such disorder in numerical simulations by using larger supercells. The overarching strategy is to (i) define an extended supercell of size $N \times N$ times the original moir\'{e} unit cell, (ii) assign electron occupations within this large cell according to a specified distribution, and (iii) compute the resulting form factors of the potential to solve the exciton Hamiltonian. In practice, we generate five distinct samples of disorder using this procedure and then take the average of their results.

\subsection{Large-cell construction}
\label{sec:appendix_disorder_A}

When the moir\'{e} lattice is perfectly filled at $\nu=1$, the real-space lattice vectors can be denoted by 
\[
   \mathbf{R}_{1} = a (1,\,0), 
   \quad
   \mathbf{R}_{2} = \frac{a}{2}\,\bigl(1,\,\sqrt{3}\bigr),
\]
and their reciprocal counterparts by 
\[
   \mathbf{G}_{1} = \frac{2\pi}{a}\,\Bigl(1,\,-\frac{1}{\sqrt{3}}\Bigr),
   \quad
   \mathbf{G}_{2} = \frac{4\pi}{a}\,\Bigl(0,\,\frac{1}{\sqrt{3}}\Bigr).
\]
{
$a$ is the moir\'{e} lattice constant, which in our modeled WSe$_2$/WS$_2$ heterobilayers arises from a near-zero twist angle and the $4\%$ lattice mismatch ($a = 8$\,nm) ~\cite{Xu2020}.}
To handle partial fillings or disordered distributions, we define an extended supercell of size $N\times N$, whose real-space vectors become 
\begin{equation}
    \mathbf{R}^{(\text{large})}_1 = N\,\mathbf{R}_1, 
    \quad
    \mathbf{R}^{(\text{large})}_2 = N\,\mathbf{R}_2.
    \label{eq:larger_vectors}
\end{equation}
The corresponding reciprocal vectors scale to 
\begin{equation}
    \mathbf{G}^{(\text{large})}_1 
    = \frac{1}{N}\,\mathbf{G}_1, 
    \quad
    \mathbf{G}^{(\text{large})}_2 
    = \frac{1}{N}\,\mathbf{G}_2.
    \label{eq:large_reciprocal}
\end{equation}

Within this enlarged real-space cell, we place electrons on discrete lattice sites according to some distribution. As an example, consider $\nu = 1/2$ and $N=12$: the total number of moir\'{e}-sites inside one extended supercell is $144$, so half-filling implies $72$ electrons assigned to the moir\'{e} sites. Here, the electron positions 
\(
   \mathbf{R}_{m} 
   = \alpha_{m}\,\mathbf{R}_1^{(\text{large})}
   + \beta_{m}\,\mathbf{R}_2^{(\text{large})}
\)
are chosen so that $\{\alpha_{m}, \beta_{m}\}$ run over discrete $N\times N$ grid points.

Once the large-cell coordinates for the occupied sites are known, we compute the moir\'{e} potential by summing over all electron positions. In reciprocal space, the \emph{form factor} of the charge distribution is
\begin{equation}
    F_{\nu}(\mathbf{G}) 
    \;=\;
    \sum_{m=1}^{M}\,
    e^{-\,i\,\mathbf{G}\cdot \mathbf{R}_{m}},
    \label{eq:form_factor_random}
\end{equation}
where $M$ is the total number of occupied sites inside the extended cell (e.g., $M=72$ for $\nu=1/2$).  Because these sites are selected based on defect density or temperature, $F_{\nu}(\mathbf{G})$ generally acquires a random phase; it does not exhibit the strong constructive interference that arises when charges form an ordered pattern. This approach can be applied to a unit cell defined at any filling factor.

\subsection{Controlling partially displaced electrons}
\label{sec:appendix_disorder_B}
Using the large-cell approach from Appendix~\ref{sec:appendix_disorder_A} to arbitrary filling $\nu$, we first define a minimal repeating unit cell that fully captures the charge-ordered pattern for that particular $\nu$. For example, if we consider $\nu = \frac{1}{2}$, the minimal repeating unit cell contains two electrons per moir\'e unit cell. We then replicate this minimal cell into an $N\times N$ supercell (with $N=12$ in this work), producing a large-cell configuration with $M$ electrons. We use $N_G = 13$ to achieve converged results. Then, we choose a certain number $\Delta$ of electrons to be displaced from their original sites. These displaced electrons are re-assigned to random sites in the extended cell, while the remaining sites remain fixed.  By varying $\Delta$ from $1$ to $M$ (the total number of electrons), we can smoothly probe the crossover between a nearly perfect charge-ordered lattice ($\Delta=1$) and a fully random distribution ($\Delta=M$). Results using this approach are shown in Fig.~\ref{fig:moire_schematic3} of the main text, where we analyze the 1$s$ and 2$s$ exciton absorption under different degrees of quasi-order. In practice, for each value of $\Delta$, we generate five distinct random configurations of displaced electrons, compute the oscillator strength for each, and then plot the averaged result in Fig.~\ref{fig:moire_schematic3}. As $\Delta$ grows, the characteristic redshifts and oscillator-strength variations (associated with long-range charge order) are partially or completely suppressed.

\subsection{Thermal fluctuations}
\label{sec:appendix_disorder_C}
In addition to static disorder, we also study how thermal fluctuations give effects on exciton states. The simulations again use large $N\times N$ supercells constructed from the unit cell of $\nu$=1, but now the electron distribution is determined by a separate classical Monte~Carlo procedure (we will cover the details in Appendix~\ref{sec:appendix_thermal}). We follow steps below:
\begin{enumerate}
    \item We fix a temperature~$T$ and specify an electron filling~$\nu$.
    \item We repeatedly sample random trial configurations of $M=\nu N^2$ electrons (similar to the above procedure) but accept or reject these trial moves based on standard Metropolis updates for the Coulomb energy.
    \item After sufficient equilibration, we acquire an ensemble of electron distributions representing the thermal state at temperature~$T$.
\end{enumerate}
Each configuration in this Monte Carlo ensemble is then used to compute $F_{\nu}(\mathbf{G})$ [Eq.~\eqref{eq:form_factor_random}], from which the exciton Hamiltonian matrix can be constructed and diagonalized.  By averaging over five Monte Carlo samples, we obtain the temperature-dependent absorption or reflectance spectra.  The main text (Sec.~\ref{subsec:thermal}) shows that this procedure captures the gradual disappearance of excitonic signatures as $T$ increases, reflecting partial or complete melting of the correlated Wigner-like state.

\section{Classical Monte Carlo simulation}
\label{sec:appendix_thermal}

In this Appendix, we detail the finite-temperature Monte Carlo method used to model thermally disordered electron configurations in a moir\'{e} superlattice.  We closely follow the approach of Ref.~\cite{Xu2020,Matty2022}, where a classical Hamiltonian describing electrons located between two parallel dielectric gates was studied via Classical Monte Carlo simulations.

\subsection{Dual gate hamiltonian}
\label{subsec:mc_method}
When a TMD is placed between two parallel metallic gates separated by distance $d=10a$ ($a = 8$\,nm), the long-range Coulomb interactions are effectively modified by the presence of image charges.  In the strong-coupling regime, each moir\'{e} site $\mathbf{r}_i$ can be either occupied ($\rho(\mathbf{r}_i)=1$) or empty ($\rho(\mathbf{r}_i)=0$).  Denoting the total number of electrons by $M = \nu \times (\text{number of moir\'{e} sites})$, the interaction energy can be written as
\begin{equation}
    \label{eq:H_gates}
    H
    \;=\;
    \tfrac{1}{2}\;
    \sum_{i \neq j}\;
    \rho(\mathbf{r}_i)\,\rho(\mathbf{r}_j)\;
    U\!\bigl(\mathbf{r}_i - \mathbf{r}_j\bigr),
\end{equation}
where $U(\mathbf{r}_i - \mathbf{r}_j)$ is the effective electrostatic potential between electrons at sites $\mathbf{r}_i$ and $\mathbf{r}_j$.  In the simplest 2D Coulomb model (without gates), one might use
\[
    U_0(r)
    \;=\;
    \frac{e^2}{4\pi\,\epsilon_0\,\epsilon}\;\frac{1}{r},
\]
but with gates located a distance $d$ above and below the TMD layer, the potential acquires an infinite series of image charges leading to a summation over Bessel functions~\cite{Xu2020,Matty2022,Throckmorton2012}.

Thus, the resulting expression takes the form
\begin{equation}
    \label{eq:H_eqn2_ref}
    U(\mathbf{r}_i - \mathbf{r}_j)
    \;=\;
    \frac{e^2}{4\pi\,\epsilon_0\,\epsilon}
    \,\frac{4}{d}
    \;
    \sum_{n=0}^{\infty}\,
    K_{0}
    \!\Bigl(\,\pi\,(2n+1)\,
              \tfrac{\bigl|\mathbf{r}_i - \mathbf{r}_j\bigr|}{\,d}
         \Bigr),
\end{equation}
where $K_{0}(x)$ is the modified Bessel function of the second kind. 

We then incorporate a filling factor $\nu$ to control the number of electrons in the moir\'{e} superlattice.  At fractional fillings (e.g., $\nu=1/3,\,2/5,\,1/2,\dots$), classical simulations find crystalline (Wigner-like) charge-ordered states at low temperatures.

\subsection{Metropolis Monte Carlo sampling}
\label{subsec:mc_method}
To simulate finite-temperature behavior, we use a standard Metropolis Monte Carlo algorithm on an $N\times N$ extended supercell~\cite{mc_book, spin_Qing2011}. For each fractional filling factor (e.g., $\nu = 1/2,\,1/4,\,1/7,\,1/3,\,2/5$), we select different $N$ values so that each extended supercell faithfully reproduces an integer number of electrons per unit cell. Specifically, $\nu=1/2$ or $1/4$ uses $N=20$ (with \(N_G=20\)), $\nu=1/7$ uses $N=14$(with \(N_G=16\)), $\nu=1/3$ uses $N=12$(with \(N_G=14\)), and $\nu=2/5$ uses $N=15$ (with \(N_G=16\)). Likewise, for $1-\nu$ filling, we use the same values as those for $\nu$ (so, for instance, $\nu=1/3$ and $\nu=2/3$ share the same \(N=12\) and \(N_G=14\))). Our resulting Classical Monte Carlo simulations at these grid sizes are consistent with the phase transitions reported in Ref.~\cite{Xu2020,Matty2022}. Concretely:

\begin{enumerate}
    \item \textbf{Initialization:} 
    Distribute $M = \nu \times (N^2)$ electrons on the $N^2$ moir\'{e} sites in random initial arrangement.
    
    \item \textbf{Local moves:}
    Randomly select a site $i$ occupied by an electron and a site $j$ that is empty, and attempt to swap their occupancies.  Calculate the resulting energy change $\Delta E$ using $H$ in Eqs.~\eqref{eq:H_gates}--\eqref{eq:H_eqn2_ref}.
    
    \item \textbf{Metropolis acceptance:}
    Accept the swap with probability 
    \[
       p_{\rm accept}
       \;=\;
       \min\bigl\{1,\;
            e^{-\beta\,\Delta E}
       \bigr\},
       \quad
       \beta \equiv \tfrac{1}{k_B T}.
    \]
    
    \item \textbf{Equilibration and measurements:}
    Repeat many such updates.  After the system equilibrates, sample electron configurations $\{\mathbf{r}_i\}$ at regular intervals for statistical averaging.
    
    \item \textbf{Extract electron distribution:}
    The resulting thermal ensemble of occupation configurations $\rho(\mathbf{r}_i)$ encodes how partial melting of the charge order emerges with increasing $T$.
\end{enumerate}

\section{2$s$ and 2$p$ exciton states at different fractional fillings}
\label{sec:appendix_2s2p}

In this Appendix, we present additional data on the 2\(s\) and 2\(p\) exciton energies under various fractional fillings, illustrating how dual-gate control of the moir\'{e} device can systematically modify the splitting between 2\(s\) and 2\(p\). As discussed in the main text, such tunability is particularly relevant for THz detection, because the 2\(s\)--2\(p\) energy difference often lies in the few-meV range (\(\sim\)THz frequencies). Adjusting the filling factor via top and bottom gates can therefore shift this energy difference and enable a gate-tunable THz sensor design.

Table~\ref{tab:2s2p_energies} summarizes illustrative energies of the 2\(s\), \(2p_{1}\) (lower-energy branch), and \(2p_{2}\) (higher-energy branch) exciton states at four representative fractional fillings: \(\nu = 1/4,\;1/3,\;2/5,\;\text{and}\;1/2\). In each case, the relative energies of \(2p\)  may lie slightly below the 2\(s\) level, and the magnitude depends on the interplay of Coulomb interactions with correlated states of moir\'{e} structure.

\begin{table}[h]
    \centering
    \setlength{\tabcolsep}{12pt}
    \caption{Illustrative energies (in meV) of the 2\(s\), \(2p_{low}\) (lower-energy branch), and \(2p_{high}\) (higher-energy branch) excitons at four fractional fillings \(\nu\). }
    \label{tab:2s2p_energies}
    \renewcommand{\arraystretch}{1.2}
    \begin{tabular}{c|c|c|c}
    \toprule
    Filling \(\nu\)  & \(E_{2p_{low}}\)   & \(E_{2p_{high}}\)   & \(E_{2s}\)   \\ 
    \hline
    \(1/4\) & \(-12.02\) & \(-12.02\) & \(-8.84\) \\
    \(1/3\) & \(-11.30\) & \(-11.30\) & \(-5.96\) \\  
    \(2/5\) & \(-12.79\)  & \(-12.78\) & \(-10.77\) \\
    \(1/2\) & \(-14.23\)  & \(-10.05\)  & \(-9.82\) \\  
    \bottomrule
    \end{tabular}
\end{table}

In particular, for the \(\nu = 1/3\) filling, the energetic separation between 2\(s\) and \(2p\) can be around \(\sim\!5.3\)\,meV, placing it firmly in the THz range. Meanwhile, at \(\nu = 1/2\), our hypothetical scenario shows two distinct 2\(s\)--2\(p\) gaps of approximately 0.3\,meV and 4.4\,meV, arising from differences in how the $x$ and $y$ \(p\)-orbitals (e.g., \(2p_{low}\) and \(2p_{high}\)) interact with the nematic-like type order(Fig.~\ref{fig:wavefunctions})(d). Notably, this implies that a single device at half-filling could accommodate two separate THz-detection channels, each tuned to a different 2\(s\)--2\(p\) resonance.

\end{document}